\documentclass[twocolumn]{autart}    

\usepackage{enumitem}
\usepackage{xcolor}
\usepackage{amsmath}
\usepackage{graphicx} 
\usepackage{subcaption}
\usepackage{amsfonts}
\let\theoremstyle\relax
\usepackage{amsthm}
\newtheorem{definition}{Definition}
\newtheorem{problem}{Problem}
\newtheorem{theorem}{Theorem}
\newtheorem{lemma}{Lemma}
\newtheorem{corollary}{Corollary}
\newtheorem{proposition}{Proposition}
\newtheorem{remark}{Remark}
\usepackage{algorithm}
\usepackage{algpseudocode}%
\usepackage{wrapfig}
\usepackage{graphicx}
\begin{document}

\begin{frontmatter}

\title{Optimal Sensor and Actuator Selection for Factored Markov Decision Processes: Complexity, Approximability and Algorithms} 

\thanks[footnoteinfo]{ Corresponding author: Jayanth Bhargav}

\author[]{Jayanth Bhargav}\ead{jbhargav@purdue.edu},    
\author[]{Mahsa Ghasemi}\ead{mahsa@purdue.edu},               
\author[]{Shreyas Sundaram}\ead{sundara2@purdue.edu}  

\address{Elmore Family School of Electrical and Computer Engineering, Purdue University}  

\begin{abstract}
Factored Markov Decision Processes (fMDPs) are a class of Markov Decision Processes (MDPs) in which the states (and actions) can be factored into a set of state (and action) variables and can be encoded compactly using a factored representation. In this paper, we consider a setting where the state of the fMDP is not directly observable, and the agent relies on a set of potential sensors to gather information. We formulate the problem of selecting a set of sensors for fMDPs (under a limited budget) to maximize the infinite-horizon discounted return provided by the optimal policy. We show the fundamental result that it is NP-hard to approximate this problem to within a factor of $n^{1-c}$ for any $c > 1$, where $n$ is the number of state variables. Our inapproximability results for sensor selection also extend to a general class of Partially Observable MDPs (POMDPs). We also consider the dual problem of budgeted actuator selection (at design-time) to maximize the expected return under the optimal policy, for which we establish similar inapproximability results. {Finally, we consider a simple greedy algorithm and empirically show that, despite the lack of formal theoretical guarantees, it performs effectively in practice, achieving on average over \(70\%\) of the optimal solution value across a variety of real-world and randomly generated problem instances.}

\end{abstract}

\end{frontmatter}

\section{Introduction}
\label{sec:introduction}
\begin{center}
\begin{figure}[htbp]
\begin{center}
    \includegraphics [width = 220pt]{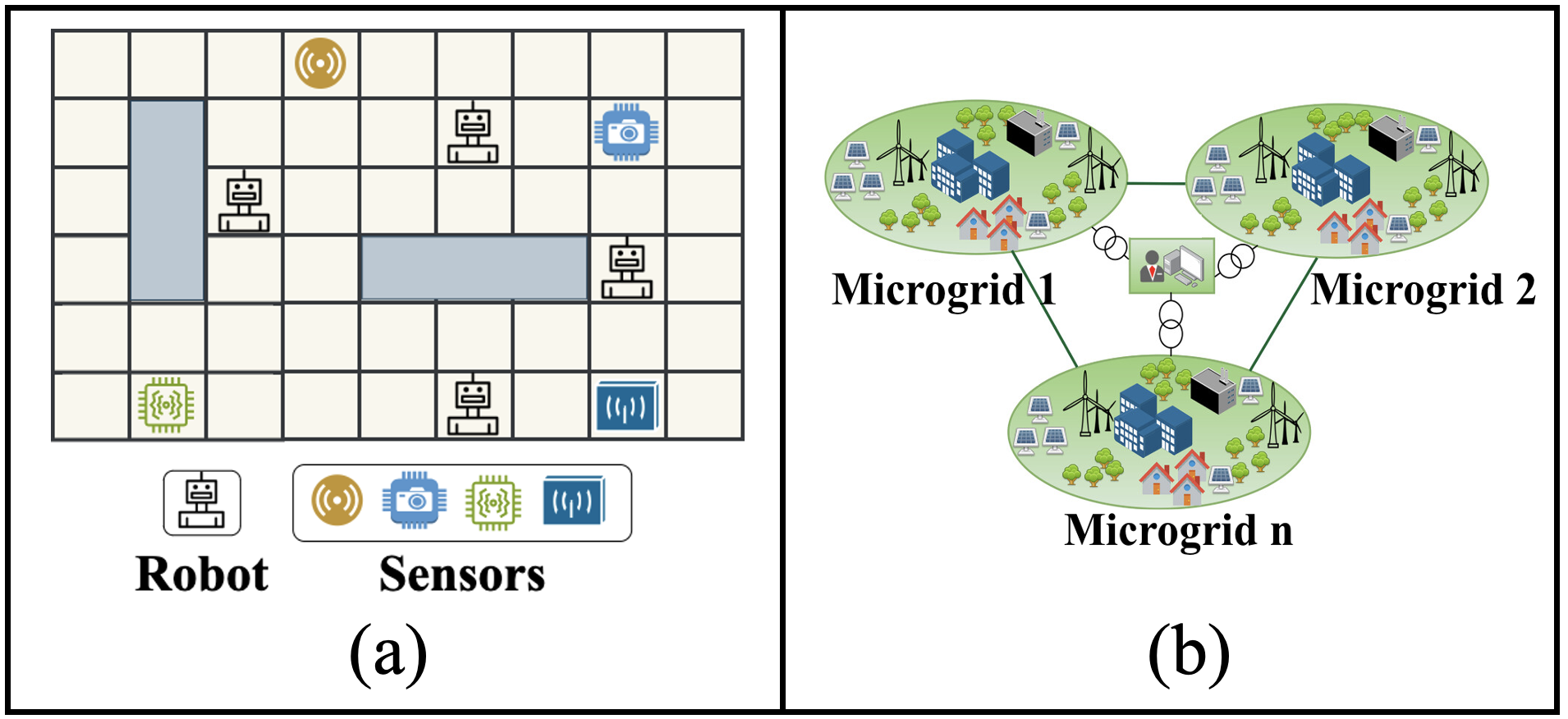}
    \caption{{(a) A multi-robot system in an environment with sensors; (b) An interconnected micro grid system.}}
    \label{fig:intro}
\end{center}
\end{figure}
\end{center}

Markov Decision Processes (MDPs) have been widely used to model systems in decision-making problems such as autonomous driving, multi-agent robotic task execution, large data center operation and machine maintenance \cite{ghasemi2019online}. In many real-world sequential decision-making problems, the state space is quite large (growing exponentially with the number of state variables). However, many large MDPs often admit significant internal structure, which can be exploited to model and represent them compactly. The idea of compactly representing a large structured MDP using a factored model was first proposed in \cite{boutilier1995exploiting}. In this framework, the state of a large MDP is factored into a set of \textit{state variables}, each taking values from their respective domains. Then, a \textit{dynamic Bayesian network (DBN)} can be used to represent the transition model. Assuming that the transition of a state variable only depends on a small subset of all the state variables, a DBN can capture this dependence compactly. Furthermore, the reward function can also be factored into a sum of rewards related to individual variables or a small subset of variables. Factored MDPs (fMDPs) exploit both \textit{additive} and \textit{context-specific} structures in large-scale systems. 
Consider the multi-agent planning problem studied in \cite{guestrin2001multiagent}, where a set of robots (denoted by $\mathcal{N}$)  must coordinate to perform tasks in an environment (see Figure \ref{fig:intro}(a)), despite uncertainty over their states. Each robot $j \in \mathcal{N}$ has its local utility $Q_j$, which depends on its own state and actions, as well as those of nearby robots. The goal is to maximize the total utility $Q = \sum_{j \in \mathcal{N}} Q_j$. Since robots have uncertainty over their states, the utility $Q_j$ of each robot is a function of its belief, which it updates using observations from sensors and shared information. A central system designer has to deploy sensors in the environment that maximize observability and improve the belief of each robot to ensure better coordination. Due to resource constraints (e.g., energy, cost), it is not feasible to equip the environment with many sensors. As a result, optimally selecting a subset of sensors within a budget that maximizes joint utility  becomes critical.
\vspace{-0cm}
\vspace{-0cm}
As another example, consider a complex electric distribution network consisting of interconnected micro-grid systems, containing generation stations, transmission lines and electric loads (see Figure \ref{fig:intro}(b)). Each node in the network is prone to electric faults (e.g., generator faults, short circuit faults). Faults in a node can cascade, and cause a catastrophic power blackout in the entire network. Due to a limited budget, a grid safety designer has to select only a subset of nodes where sensors and actuators (e.g., Phasor Measurement Units (PMUs), Islanding switches) can be installed, which can help minimize fault percolation in the network, by identifying and isolating certain critical nodes or micro-grid networks.

 The multi-agent systems described above can be modelled as MDPs, and often have exponentially large state and action spaces in practice (e.g., a large number of nodes in the network). In such cases, one can leverage the internal structure of such systems and model them as fMDPs. If one does not have complete observability of the states and must gather information using a limited number of sensors or can only influence transitions of a subset of state variables using a limited number of actuators, one is faced with the problem of selecting the optimal set of sensors (or actuators) that can result in better performance of such systems (fMDPs). 

\begin{table*}[!ht]
  \centering
  \small
 \begin{tabular}{cccc}
  \hline
  \textbf{Reference(s)} & \textbf{Problem Setting}  & \textbf{Complexity} & \textbf{Properties} \\
  \hline
    \cite{bian2006utility}& Tiered network system & P & Modularity (exact solution)  \\
    \cite{hashemi2020randomized,bhargav24a} & Linear systems, Hypothesis testing  & NP-hard & Weak-submodularity (near-optimal approx.)  \\
    \cite{ghasemi2019perception,tzoumas2015minimal}& POMDPs, Kalman filtering  &NP-hard & Submodularity (near-optimal approx.) \\
    \cite{golovin2011adaptive}& Influence maximization in networks & NP-hard & Adaptive submodularity (near-optimal approx.)\\
    \cite{ye2020resilient}& Kalman filtering in networked systems & NP-hard & Weakly NP-hard (exact solution) \\
    \textbf{This work}& Factored MDPs & NP-hard & Inapproximable (no approximation guarantees)\\
    \hline
  \end{tabular}
  \caption{Overview of complexity and properties for a selected set of sensor and actuator selection problems studied in the literature.}
  \label{tab:1}
\end{table*}

 Sensor placement/selection has widely been studied in the context of MDPs and its variants like POMDPs. In \cite{ghasemi2019online}, the authors consider active perception under a limited budget for POMDPs to selectively gather information at runtime. However, in our problem, we consider design-time sensor/actuator selection for fMDPs, where the sensor/actuator set is not allowed to change at runtime. A body of literature considers the problem of dynamic sensor scheduling for sequential decision-making tasks and model the task of sensor selection as a POMDP \cite{krishnamurthy2007structured,ji2007nonmyopic}. However, in many real-world scenarios, the set of sensors or actuators may not be allowed to change due to reasons like regulatory requirements or the need for certified reliability.  Hence, in this paper, we focus on the problem of selecting the optimal set of sensors (or actuators) a priori (at \textit{design-time}) for fMDPs.


{The problem of sensor subset selection has been extensively explored in the controls' literature, and representative studies are summarized in Table~\ref{tab:1}. As the table indicates, each formulation varies in its objective, assumptions, and structural properties, resulting in diverse algorithmic strategies and performance guarantees. Consequently, every sensor selection problem is unique and requires careful theoretical analysis to understand its computational complexity and optimality characteristics. Similarly, there are many works that have considered the problem of actuator placement for optimizing system performance under budget constraints \cite{tzoumas2015minimal}. The authors of \cite{fuchs2013actuator} consider the problem of optimal placement of High Voltage Direct Current links in a power system, which are used to stabilize a power grid from oscillations. However, these techniques use linearized state-space models, whereas in this paper, we consider fMDP models which may not necessarily have a linear structure. 
In contrast to several existing works that establish near-optimal approximation algorithms for specific formulations, we present worst-case inapproximability results for both sensor and actuator selection in fMDPs.}



\noindent \textbf{Contributions.}
 {Firstly, we show that the problem of selecting an optimal subset of sensors at design-time for a general class of  fMDPs is NP-hard, and there is no polynomial time algorithm that can approximate this problem to within a factor of $n^{1-c}$ of the optimal solution, for any $c>1$, where $n$ is the number of state variables. In other words, this inapproximability result implies that for an objective function $f(x)$ with an optimal solution $x^*$ (i.e., a maximizer), no polynomial-time algorithm can guarantee a solution $\bar{x}$ such that $f(\bar{x}) \ge n^{1-c} f(x^*).$  We would like to emphasize that computing the optimal policy for a fMDP is known to be PSPACE-hard \cite{papadimitriou1987complexity}. Our inapproximability results go beyond the complexity of computing the optimal policy. We prove that the fMDP sensor selection problem is inapproximable even when one has access to an \textit{oracle} that can compute the optimal policy for any given instance. We show that these inapproximability results also apply to a general class of POMDPs, complementing the results in \cite{krause2008optimizing}, where the authors provide near-optimal greedy algorithms for sensor selection for a special class of POMDPs with submodular value functions. Second, we show that the same inapproximability results hold for the problem of selecting an optimal subset of actuators at design-time for a general class of  fMDPs.  Finally, complementing our inapproximability results, we present empirical studies which demonstrate that greedy algorithms achieve near-optimal solutions in {\it several instances} of the problem in practice, averaging over $70\%$ of the optimal performance.}



We considered the problem of optimal sensor selection for Mixed-Observable MDPs (MOMDPs) in the conference paper \cite{bhargav2023complexity}. We proved its NP-hardness and provided insights into the lack of submodularity of the value function. However, in this paper, we present stronger inapproximability results for both sensor and actuator selection for fMDPs, of which MOMDPs are a special case.  
\section{Problem Formulation}
In this section, we formally state the sensor and actuator selection problems for fMDPs.
A general factored MDP is defined by the following tuple: $\mathcal{M} :=(\mathcal{S} = \mathcal{S}_1  \times \cdot \cdot \cdot \times \mathcal{S}_n,\mathcal{A} = \mathcal{A}_1 \times \cdot \cdot \cdot \times \mathcal{A}_m, \mathcal{T}, \mathcal{R},  \gamma)$, where $\mathcal{S}$ is the state space decomposed into finite sub-spaces $\mathcal{S}_i$, each corresponding to a state-variable $s_i$ (which takes values from $\mathcal{S}_i$), $\mathcal{A}$ is the action space decomposed into finite sub-spaces $\mathcal{A}_i$, each corresponding to an action $a_i$ (which takes values from $\mathcal{A}_i$), $\mathcal{T}$ is the probabilistic state transition model, $\mathcal{R}$ is the reward function and $\gamma \in (0,1)$ is the discount factor. The state transition model $\mathcal{T}: \mathcal{S} \times \mathcal{A} \times \mathcal{S} \rightarrow [0,1] $ captures the relationship between state transitions of the factored state variables $s_i$. The reward function $\mathcal{R}: \mathcal{S} \times \mathcal{A} \rightarrow \mathbb{R}$, is a scalar function of the factored state variables $s_i$ and actions $a_i$. In general, a fMDP does not include an observation space or a belief state, as it assumes full observability of the system state. However, in the sensor selection problem, we consider a setting where the state variables are not observable. Consequently, sensors provide observations, forming an observation space, and the agent maintains a belief over the underlying state.

\subsection{The Sensor Selection Problem}
Consider the scenario where the agent does not have observability of the state variables $s_i$ of the fMDP. The agent has to select a subset of sensors to install, which can provide information about the state. Define $\Omega = \{\omega_i \mid i = 1,2,\hdots,s\}$ to be a collection of sensors, where each sensor $\omega_i$ provides an observation $o_i \in \mathcal{O}_i$, where $\mathcal{O}_i$ is the observation space of the sensor $\omega_i$. Each sensor $\omega_i$ has a likelihood function conditioned on the state $s \in \mathcal{S}$ and the sensor model could incorporate noise. Let $c_i \in \mathbb{R}_{\geq 0}$ be the cost we pay to select the sensor $\omega_i$, and let $C \in \mathbb{R}_{> 0}$ denote the total budget for the sensor placement. Let $\Gamma \subseteq \Omega$ be the subset of sensors selected (at design-time) that generates observations $y_{\Gamma}(t) = \{o_i(t) \mid \omega_i \in \Gamma \}$. 
 At time $t$, the agent has the following information: observations $Y^{\Gamma}_t = \{y_\Gamma (0), y_\Gamma (1), y_\Gamma (2), \cdot \cdot \cdot, y_\Gamma (t)\} $, joint actions $A_t = \{a(0), a(1), a(2), \cdot \cdot \cdot, a(t-1)\}$ and rewards $R_t = \{r(0), r(1), r(2), \cdot \cdot \cdot, r(t-1)\}$. \\
 
 \begin{remark}
 \label{remark:pomdp}
     We would like to emphasize the equivalence of this setting to a POMDP. Since the system's true state is not fully observable, an agent can only access partial observations that provide information about the underlying state. However, the key difference in our case lies in the factored state and action spaces, which is not typically present in traditional POMDPs. A fMDP where the state variables are not observable is a POMDP, and a POMDP can be expressed as a fMDP (with partial observability) with a single state variable, resulting in a trivial state space factorization. 
 \end{remark}
 Akin to POMDPs, the agent maintains a belief over the states of the fMDP, $b \in \mathcal{B}$, where $\mathcal{B}$ is the belief space, which is the set of probability distributions over the states in $\mathcal{S}$. The agent updates this belief based on the observation likelihood and state transition functions using a belief update rule (e.g., Bayes' rule). In this case, the expected reward is \textit{belief-based}, denoted as $\rho(b,a)$, and is given by $\rho(b,a) = \sum_{s} b(s)r(s,a)$,  where $b(s)$ is the belief over the state $s$, $a$ is the action and $r(s,a)$ is the reward obtained for taking action $a$ in the state $s$. 
For any sensor set $\Gamma \subseteq \Omega $, let $\mathcal{H}_t = \{ Y^{\Gamma}_t,A_t,R_t\}$ denote the set of all the information the agent has until time $t$. Define $\Pi_\Gamma = \{ \pi_\Gamma \mid \pi_\Gamma : \mathcal{H}_t \rightarrow \mathcal{A}\}$ to be a class of history-dependent  policies that map from a set containing all the information known to the agent until time $t$ to the action $a_t$ which the agent takes at time $t$.  The agent seeks to maximize the expected infinite-horizon return, given the initial belief $b_0$, by finding an optimal policy satisfying  $\pi_\Gamma^*=$ $\arg \max _{\pi \in \Pi_\Gamma} V^\pi\left(b_0\right)$ with $V^\pi\left(b_0\right)=\mathbb{E}\left[\sum_{t=0}^{\infty}\gamma^t \rho_t \mid b_0, \pi\right]$,
where $\rho_t$ is the expected reward obtained at time $t$. 
The value function $V^\pi_{n}(b)$ under a policy $\pi$ can be computed using value iteration with $V^{\pi}_0(b)=0$, and $V^\pi(b)= \lim_{n \rightarrow \infty} V^{\pi}_{n}(b)$ \cite{araya2010closer}. 
 The goal is to find an optimal subset of sensors $\Gamma^* \subseteq \Omega$, under the budget constraint, that can maximize the expected value $V^*_\Gamma$ under the optimal policy. Specifically, we aim to solve the optimization problem: 
\begin{align}
& \hspace{10pt}  \max_{\Gamma \subseteq \Omega} \hspace{5pt}  V_\Gamma^* ; \quad\textit{s.t.} \sum_{\omega_i \in \Gamma} c_i  \leq C.
\end{align}
\vspace{-0.2mm}
In order to analyze the complexity of the above problem, we define its decision version as  the \textit{Factored Markov Decision Process Sensor Selection (fMDP-SS)  Problem}.

\begin{problem}[fMDP-SS Problem]
\label{prob:momdpss}
Consider a fMDP $\mathcal{M}$ and a set of sensors $\Omega$, where each sensor $\omega_i \in \Omega$ is associated with a cost $c_i \in \mathbb{R}_{\geq 0}$. For a value $V \in \mathbb{R}$ and sensor budget $C \in \mathbb{R}_{> 0}$, is there a subset of sensors $\Gamma \subseteq \Omega$, such that the expected return $V_\Gamma^*$ for the optimal policy in $ \Pi_\Gamma$ satisfies  $V_\Gamma^* \geq V$ and the total cost of the sensors satisfies $\sum_{\omega_i \in \Gamma} c_i \leq C$?
\end{problem}

\subsection{The Actuator Selection Problem}
Consider the scenario where there are no actuators installed initially, and the agent cannot influence the transitions of state variables $s_i$ of the fMDP. However, the agent has complete observability of the state variables $s_i$. In this case, the agent has to select a subset of actuators to install. 
Define $\Phi = \{\phi_i \mid i = 1,2,\hdots,a\}$ to be a set of actuators, where the actuator $\phi_i$ provides a set of actions $\mathcal{A}_i$, and action $a_i \in \mathcal{A}_i$ can influence the state transitions of some or all of the state variables in $s$. Let $k_i \in \mathbb{R}_{\geq 0}$ be the cost we pay to install the actuator $\phi_i$, and let $K \in \mathbb{R}_{> 0}$ denote the placement budget. Let $\Upsilon \subseteq \Phi$ be the subset of actuators selected. Let $\mathcal{A}_{\Upsilon} = \Pi_{\phi_i \in \Upsilon} \mathcal{A}_i = \{ \{{a}_i\}| \phi_i \in \Upsilon \} $ denote the joint action space available to the agent by selecting the actuator set $\Upsilon$. If no actuator is selected, it is equivalent to having a default action $a_{d}$ with which the agent stays in its current state \textit{w.p.} 1. The rewards depend on the joint state-action pairs, i.e., $r(s,\Bar{a})$ is the reward for taking the joint action $\Bar{a}$ in state $s$. Define $\Pi_\Upsilon = \{ \pi_\Upsilon \mid \pi_\Upsilon : \mathcal{H}_t \rightarrow \mathcal{A}_\Upsilon\}$ to be a class of history-dependent  policies that map from a set containing all the information $\mathcal{H}_t$ known to the agent until time $t$  to action $\Bar{a}_t \in \mathcal{A}_\Upsilon$ which the agent takes at time $t$. The goal of the agent is to choose the best actuator set $\Upsilon^* \subseteq \Phi$ which can maximize the expected  return under the optimal policy of the resulting MDP. {This  return is computed for the optimal policy satisfying  $\pi^*=\arg \max_{\pi \in \Pi_\Upsilon} V^\pi$ with $V^\pi=\mathbb{E}\left[\sum_{t=0}^{\infty}\gamma^t r_t \mid \pi, s_0 \right]$,
where $r_t$ is the reward obtained at time $t$ and $s_0 \in \mathcal{S}$ is the given initial state.} We note that the optimal policy in this case is stationary and deterministic.  Let $V_{\Upsilon}^*$ denote the expected return under the optimal policy for the actuator set $\Upsilon$. We aim to solve the following: 
\begin{align}
& \hspace{10pt}  \max_{\Upsilon \subseteq \Phi} \hspace{5pt}  V_\Upsilon^*; 
 \quad \textit{s.t.} \sum_{\phi_i \in \Upsilon} k_i  \leq K.
\end{align}
Similar to the sensor selection problem (fMDP-SS), we define the decision version of the actuator selection problem above as the \textit{Factored Markov Decision Process Actuator Selection Problem (fMDP-AS Problem)}. \\

\begin{problem}[fMDP-AS Problem]
\label{prob:momdpas}
Consider a fMDP $\mathcal{M}$ and a set of actuators $\Phi$, where each actuator $\phi_i \in \Phi$ is associated with a cost $k_i \in \mathbb{R}_{\geq 0}$. For a value $V \in \mathbb{R}$ and actuator budget $K \in \mathbb{R}_{> 0}$, is there a subset of actuators $\Upsilon \subseteq \Phi$, such that the expected return $V_\Upsilon^*$ for the optimal policy in $ \Pi_\Upsilon$ satisfies  $V_\Upsilon^* \geq V$ and the total cost of the actuators selected satisfies $\sum_{\phi_i \in \Upsilon} k_i \leq K$?
\end{problem}

\section{Complexity and Approximability Analysis}

In this section, we present the inapproximability results for fMDP-SS and fMDP-AS problems. We begin by introducing the Set Cover Problem, a classical problem in combinatorics that serves as the foundation for many hardness results in optimization. The NP-hardness and inapproximability results established in this work are derived through a reduction from this problem. {For a brief overview of the relevant concepts from complexity theory, see Appendix \ref{app:complexity_theory}.}

\subsection{Set Cover Problem}
The \textit{Set Cover Problem} is a classical problem in combinatorics and complexity theory. Given a set of $n$ elements $\mathcal{U} = \{u_1,u_2, \hdots, u_n \}$ called the universe and a collection of $m$ subsets $\mathbb{S} = \{ S_1, S_2, \hdots, S_m | S_i \subseteq \mathcal{U}\}$, where each subset $S_i$ is associated with a cost $c(S_i) \in \mathbb{R}_{\geq 0}$ and  the union of these subsets equals the universe $\mathcal{U} = S_1 \cup S_2 \cup \cdots \cup S_m$, the set cover problem is to identify the collection of subsets in $\mathbb{S}$ with minimum cost, whose union contains all the elements of the universe. Let $\mathbb{S}_c$ denote the collection of subsets selected.  We wish to solve the following optimization problem:
\begin{align*}
& \hspace{1pt}  \min_{\mathbb{S}_c \subseteq \mathbb{S}} \hspace{5pt}  \sum_{S_i \in \mathbb{S}_c} c(S_i) ; \quad \textit{s.t.}  \hspace{0pt} \bigcup_{S_i \in \mathbb{S}_c} S_i = \mathcal{U}.
\end{align*}
In order to establish NP-hardness, we now define the decision version of this problem under uniform set selection costs, referred to as the \textsc{SetCover} Problem.
\\
\begin{problem}[\textsc{SetCover} Problem]
     Consider a universal set of $n$ elements $\mathcal{U}:= \{ u_1, u_2, \hdots, u_n \}$ and a collection of its subsets $\mathbb{S}:= \{ S_1, S_2, \hdots, S_m \}$. For a positive integer $k$, is there a collection $\mathbb{S}_k$ of at most $k$ subsets $S_i$ in $\mathbb{S}$ such that $\bigcup_{S_i \in \mathbb{S}_k} S_i = \mathcal{U}$?
 \end{problem} \vspace{3pt}
The \textsc{SetCover} Problem is NP-Complete \cite{karp2010reducibility}. {We now begin by establishing the complexity of the fMDP-SS problem. To this end, we consider a single-state MDP as defined in Example 1 below and show in Lemma~\ref{lma1} that when a sensor is installed (i.e., the state is observable), the optimal policy yields a positive reward, whereas when no sensor is installed (i.e., the state is unobservable), the optimal policy yields zero return. This example serves as the foundational building block for the inapproximability result presented later in Theorem~\ref{thm1}. All omitted proofs can be found in Appendix \ref{app:proofs}.}

\begin{center}
\begin{figure}[htbp]
\begin{center}
    \includegraphics[width = 180pt]{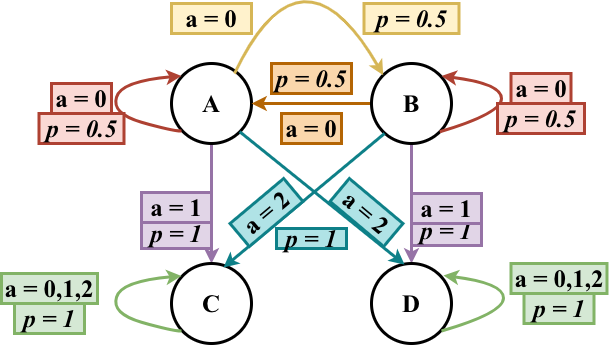}
    \caption{{State transition diagram of $\Bar{\mathcal{M}}$.}}
    \label{fig:small_mdp}
\end{center}
\end{figure}
\end{center}

\textit{Example 1:} Consider an MDP $\Bar{\mathcal{M}}:= \{\mathcal{S},\mathcal{A},  \mathcal{T}, \mathcal{R}, \gamma, b_0\}$ with state space $\mathcal{S} = \{ A, B, C, D\}$, actions $\mathcal{A} = \{ 0,1,2\}$, transition function $\mathcal{T} :   
   \Biggl\{ \mathcal{T}_{0} = \left[\begin{smallmatrix}
  0.5 & 0.5 & 0 & 0\\ 
 0.5 & 0.5 & 0 & 0 \\
 0 & 0 & 1 & 0 \\
 0 & 0 & 0 & 1 
\end{smallmatrix}\right] ; \mathcal{T}_1 = \left[\begin{smallmatrix}
  0 & 0 & 1 & 0\\ 
 0 & 0 & 0 & 1 \\
 0 & 0 & 1 & 0 \\
 0 & 0 & 0 & 1 
\end{smallmatrix}\right] ; \mathcal{T}_2 = \left[\begin{smallmatrix}
  0 & 0 & 0 & 1\\ 
 0 & 0 & 1 & 0 \\
 0 & 0 & 1 & 0 \\
 0 & 0 & 0 & 1 
\end{smallmatrix}\right]\Biggr\}$ for $a = 0$, $a = 1$ and $a = 2$, respectively, reward function $
\mathcal{R}(s,a) = (
      r(A,0) = 0, r(B,0) = 0,
      r(A,1)=R ,r(B,1)=-(1+\delta)R, r(A,2)=-(1+\delta)R ,r(B,2)= R,  r(C,\cdot)= R, r(D,\cdot)=- (1+\delta)R)$ with $ R,\delta >0$, discount factor $\gamma \in (0,1)$ and initial distribution $b_0 = [0.5,0.5,0,0]$. Fig. \ref{fig:small_mdp} describes state-action transitions along with their probabilities. 
The state space of this MDP corresponds to only one state variable $s$ and the agent can measure this by selecting a noiseless sensor $\omega = s$. The initial state of the MDP at $t = 0$ is either $s_0 = A$ or $s_0 = B$, with equal probability.  
\vspace{1cm}
 \begin{lemma}
 \label{lma1}
For the MDP $\Bar{\mathcal{M}}$ defined in Example 1, the following holds for any $\gamma\in (0,1)$:
\begin{itemize}
    \item [(i)] If the state of  $\Bar{\mathcal{M}}$ is measured (or observable) using the sensor $\omega$, the infinite-horizon expected return under the optimal policy is $V^*(s) = \frac{R}{(1-\gamma)}$ and the optimal policy ensures that the state of the MDP is $s_t=C$, for all $t>0$. 
    \item[(ii)] If the state of $\Bar{\mathcal{M}}$ is not measured i.e., there is no sensor installed and the agent only has access to the sequence of actions and rewards, but not the current state $s$, then the infinite-horizon expected reward beginning at belief $b_0$, under the optimal policy is $V^*(b) = 0$ and the optimal policy ensures that the state of the MDP is $s_t\notin  \{C,D\}$, for all $t>0$. 
\end{itemize}
\end{lemma}

We are now in place to provide the following result characterizing the complexity of the fMDP-SS problem.\\

\begin{theorem} \label{thm1} Unless $P=NP$, there is no polynomial-time algorithm that can approximate the fMDP-SS Problem to within a factor of $n^{1-c}$, for any $c > 1$.
\end{theorem}
\begin{proof}
    We consider an instance of the \textsc{SetCover} Problem with a collection of $m$ sets over $n$ elements of the universe and reduce it to an instance of the fMDP-SS Problem, similar to the reduction presented in \cite{kempe2003maximizing} and \cite{schoenebeck2019beyond}. We construct an fMDP consisting of $N$ identical MDPs $\mathcal{M}^* := \{ \mathcal{M}_1, \mathcal{M}_2, \hdots, \mathcal{M}_N \}$, where $N = m+ n +n^c$, for some large $c>1$ (see Figure \ref{fig:setcovermdp}). Each MDP $\mathcal{M}_i$ is similar to the MDP defined in Example 1, except for the reward function, which we will define below. The state of fMDP $\mathcal{M}^*$ has $N$ state variables and the joint action consists of $N$ actions, each corresponding to an MDP $\mathcal{M}_i$ and can be defined as the following tuple: $\mathcal{M}^* :=(\mathcal{S},\mathcal{A}, \mathcal{T}, \mathcal{R},  \gamma)$. The states of the MDPs are independent of each other, however the reward function for each MDP is a function of its own state as well as the states of the other MDPs. We will now explicitly define the state space, action space, transition function, reward function and discount factor as follows. \\
    \textit{State Space $\mathcal{S}$}: The state space $\mathcal{S}_i$ corresponds to the MDP $\mathcal{M}_i$ as defined in Example 1. We define the state space $\mathcal{S}$ of the $N$-state fMDP $\mathcal{M}^*$ as
$
 \mathcal{S} := \mathcal{S}_1 \times \mathcal{S}_2 \times \hdots \mathcal{S}_N. $
The state variable $s_i$ corresponding to the MDP $\mathcal{M}_i$ can be one-hot encoded in $4$ bits, i.e., $A = 1000, B = 0100, C = 0010$ and $D = 0001$. The complete state of the fMDP $\mathcal{M}^*$ can be represented in $4N$ bits, where $N=m+n+n^c$.\\
\textit{Action Space $\mathcal{A}$}: The  action space $ \mathcal{A}_i$ corresponds to the MDP $\mathcal{M}_i$ as defined in Example 1. We define the action space $\mathcal{A}$ of the fMDP $\mathcal{M}^*$ as 
   $\mathcal{A} := \mathcal{A}_1 \times \mathcal{A}_2 \times \hdots \mathcal{A}_N.$\\
\textit{Transition Function $\mathcal{T}$}: The overall transition function of the fMDP $\mathcal{M}^*$ is defined by a collection of $N$ transition functions, $\mathcal{T}:= \{ \mathcal{T}_1, \hdots, \mathcal{T}_N \}$, where $\mathcal{T}_i$ is the transition function of MDP $\mathcal{M}_i$ as described in Example 1. The state transition probability of the fMDP can be computed as follows: 
$ \mathcal{T}(\mathbf{s'} | \mathbf{s}, \mathbf{a} ) = \prod_{i = 1}^{N} \mathcal{T}_i(s_i' | s_i, a_i )$,
 where $\mathbf{s} = [s_1, \hdots, s_N]$ is the joint state, $\mathbf{a} = [a_1, \hdots, a_N]$ is the joint action and $\mathcal{T}_i(s_i' | s_i, a_i )$ is as defined in the transition function of the $i$'th state variable with respect to the $i$'th action variable. Each local factor $\mathcal{T}_i$ depends only on its corresponding state--action pair $(s_i, a_i)$ 
and can be represented by a constant-size conditional probability table. 
Since there are $N = \text{poly}(m,n)$ such factors, the overall transition model requires only 
$\text{poly}(m,n)$ bits.\\
\textit{Discount Factor $\gamma$}: Let the discount factors $\gamma_i$ 's of all MDP's $\mathcal{M}_i$ be equal to each other, $ \gamma_1 = \hdots = \gamma_N = \gamma.$\\
\textit{Reward Function $\mathcal{R}$}: 
We first define the structure of the fMDP which captures the influence that the states of individual MDPs have over the reward functions of the other MDPs. The reward functions of the MDPs $\{ \mathcal{M}_1, \hdots, \mathcal{M}_m \}$  are independent of each other, and are defined in Example 1. For MDP $\mathcal{M}_i$, where $i = m+1, \hdots, m+n$, the reward function $\mathcal{R}_i$ is a function of the states of the first $m$ MDPs, i.e., $\hat{s} = [s_1, \hdots, s_m]$. Define $\Tilde{s}_i =  (\vee_{j: u_i \in S_j} s_j) \wedge (0010)$ for $m+1 \leq i \leq m+n$, where $\vee$ is a bit-wise Boolean OR operation \footnote{The bit-wise Boolean  OR operation over two n-bit Boolean strings X and Y is a n-bit Boolean string Z, where the $i$'th bit of Z is obtained by applying the Boolean OR operation to the $i$'th bit of X and $i$'th bit of Y.} and $\wedge$ is a  bit-wise Boolean AND operation \footnote{The bit-wise Boolean AND operation over two n-bit Boolean strings X and Y is a n-bit Boolean string Z, where the $i$'th bit of Z is obtained by applying the Boolean AND operation to the $i$'th bit of X and $i$'th bit of Y.} over the states. The reward function $\mathcal{R}_i$ for $i = m+1, \hdots, m+n$ is given by
$\mathcal{R}_i(\hat{s})= 
        R$ if $\Tilde{s}_i = 0010$ or 
       $\mathcal{R}_i(\hat{s})= 0$, otherwise. 
The above reward function means that the reward of an MDP $\mathcal{M}_i$, where $i = m+1, \hdots, m+n$ is $R$ if the state $s_j$ of MDP $\mathcal{M}_j$ is $C$ (or $0010$) for any $j$ such that $u_i \in S_j$ in the \textsc{SetCover} instance. 
 The reward function $\mathcal{R}_i$ for $m+n+1 \leq i \leq m+n+n^c$ is given by
\begin{equation}
\mathcal{R}_i(\Hat{s})=
    \begin{cases}
        R & \text{if } \Tilde{s}_{m+1} \wedge, \hdots, \wedge \Tilde{s}_{m+n} = 0010 \\
        0 & \text{otherwise }
    \end{cases}.
\end{equation}
According to the above reward function, the reward of an MDP $\mathcal{M}_i$ for $m+n+1 \leq i \leq m+n+n^c$ is $R$ only if all $\Tilde{s}_i$'s are equal to $0010$. The overall reward function of the fMDP $\mathcal{M}^*$ is given by
$\mathcal{R} := \sum_{i = 1}^{N} \mathcal{R}_i.$
Figure \ref{fig:setcovermdp} shows the dependence of the rewards of each MDP in the fMDP on the states of all the MDPs derived using the \textsc{SetCover} instance. The rewards of MDPs in Layer 1 are independent of each other. 
\begin{figure*}[!ht]
\centering
  \includegraphics[width=0.6\textwidth]{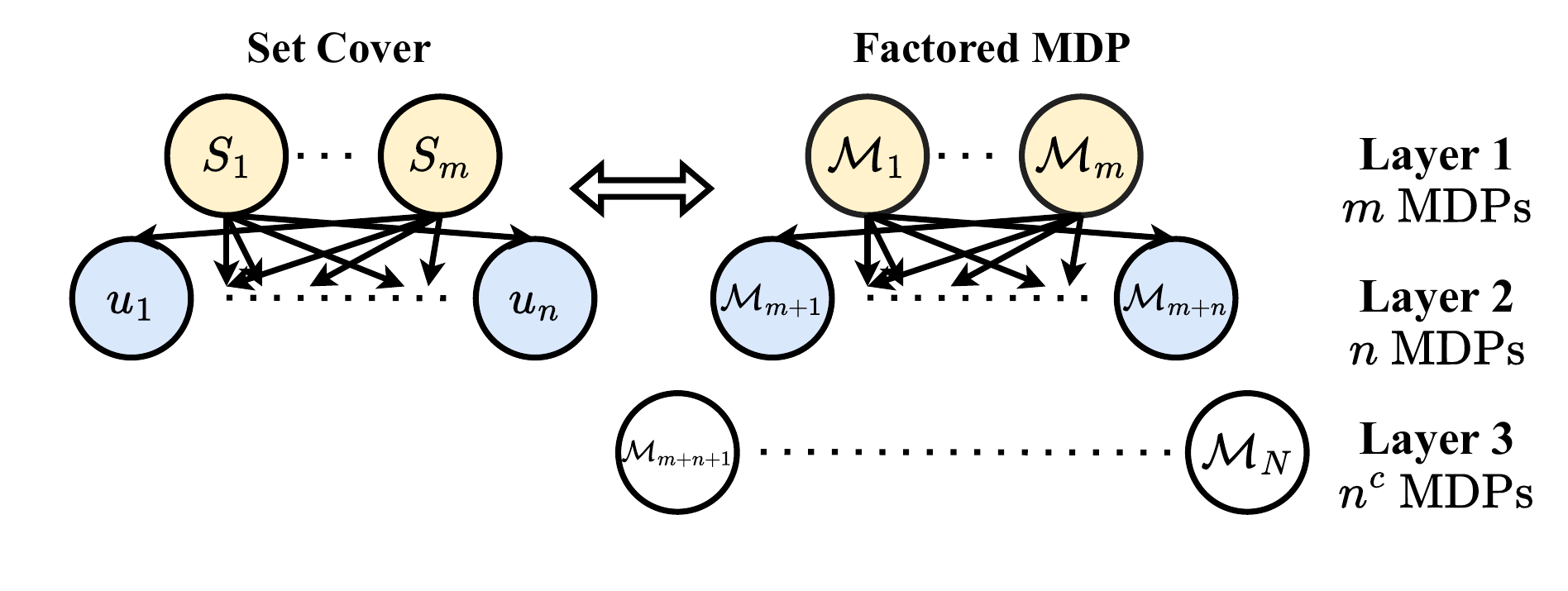}
\caption{{Reduction from \textsc{SetCover} to fMDP-SS/ fMDP-AS: MDPs in Layer 1 correspond to $m$ subsets and MDPs in Layer 2 correspond to the $n$ elements in the \textsc{SetCover} problem. The reward of an MDP in Layer 1 depends on its own states and actions. The rewards of the MDPs in Layers 2 and 3 depend on the states of all the MDPs in Layer 1.} }
\label{fig:setcovermdp}
\end{figure*}
The reward of MDPs in Layers 2 and 3 depend on the states of all MDPs in Layer 1. Note that the reward function takes as an input the joint state $\mathbf{s}$ of the fMDP and computes the reward using bit-wise Boolean operations over at most $4N$ bits, the complexity of which is $poly(m,n)$. {The parameters $R \in \mathbb{R}$ and $\gamma \in (0,1)$ are fixed constants, independent of the problem size $n$, and therefore admit constant-size bit representations. For instance, we set $R = 1$ and $\gamma = 0.9$. This ensures that they have a constant encoding size, i.e.,  $\mathcal{O}(1)$, preserving the polynomial size of the encoding.
}

Let the sensor budget be $C = k$, where k is the maximum number of sets one can select in the \textsc{SetCover} Problem. Let $\Omega$ denote the set of sensors, where each sensor $\omega_i \in \Omega$ has a selection cost $c_i$, and corresponds to the MDP $\mathcal{M}_i$, where $1 \leq i \leq N$. Let  $c_i = 1$  for $1 \leq i \leq m$ and $c_i = k+1$ for $ m+1 \leq i \leq N$.  Let the value function threshold $V$ for the fMDP be $V = (k + \gamma n + \gamma n^c) R/(1-\gamma)$. Note that for the specified $R$ and $\gamma$ values, $V$ can be encoded in $\mathcal{O}(n)$ bits.  Additionally, for the specified sensor budget and sensor costs, only a subset of sensors corresponding to MDPs $\{ \mathcal{M}_1, \hdots, \mathcal{M}_m\}$ can be selected. 

 We now have an instance of the fMDP-SS Problem, obtained by reducing an instance of the \textsc{SetCover} Problem. If the answer to the \textsc{SetCover} Problem is True, there is a full set cover $\mathbb{S}_k$ of size $k$ which satisfies $\bigcup_{S_i \in \mathbb{S}_k} S_i = \mathcal{U}$. By deploying sensors on $k$ MDPs $\mathcal{M}_i$ corresponding to sets $S_i$ in $\mathbb{S}_k$, where $1 \leq i \leq m$, we have from Lemma \ref{lma1} that these MDPs will be in state $C$ for $t \geq 1$ and have an infinite-horizon expected return of $R/(1-\gamma)$ under the optimal policy. For $t\geq 1$, $\Tilde{s}_i$ evaluates to $0010$. Thus, according to the specified reward function, each MDP $\mathcal{M}_i$ for $m+1 \leq i \leq m+n$ has an infinite-horizon expected return of $\gamma R/(1-\gamma)$. It also follows that each MDP $\mathcal{M}_i$ for $m+n+1 \leq i \leq N$ has an infinite-horizon expected return of $\gamma R/(1-\gamma)$. The total infinite horizon expected return of the fMDP is  $V_{\Gamma}^{*(1)} = (k+\gamma n+ \gamma n^c) R/(1-\gamma)$. Thus, the answer to the fMDP-SS instance is also True. 
  
  Conversely, if the answer to fMDP-SS Problem is True, there is a sensor set $\Gamma$ with $\sum_{\omega_i \in \Gamma} c_i \leq k$ sensors, installed on at most $k$ MDPs $\mathcal{M}_i$ where $1 \leq i \leq m$ and $V_{\Gamma}^* \geq (k + \gamma n + \gamma n^c) R/(1-\gamma)$. It follows from the specified reward function and the infinite-horizon return obtained in the fMDP that the collection of subsets $S_i \in \mathbb{S}$ which correspond to the MDPs $\mathcal{M}_i$ in the fMDP on which sensors are installed, collectively cover all the elements of the universe $\mathcal{U}$ in the \textsc{SetCover} instance. Thus, the answer to the \textsc{SetCover} instance is True.

  If the answer to \textsc{SetCover} Problem is False, then there is no set cover of size $k$ that covers all the elements of the universe $\mathcal{U}$. In this case, the maximum number of elements that can be covered by any $k$ Set Cover is $n-1$. This means that, by deploying sensors on the corresponding $k$ MDPs, at most $k$ MDPs in Layer 1 can have an expected infinite-horizon return of $R/(1-\gamma)$ and at most $n-1$ MDPs in Layer 2 can have an expected infinite-horizon return of $\gamma R/(1-\gamma)$. The maximum value of the expected return would be $V_{\Gamma}^{*(2)} = (k+ \gamma(n-1)) R/(1-\gamma)$. Define the ratio $r_{approx}$ as
  \begin{equation} \label{eq:badratio}
      r_{approx} = \frac{V_{\Gamma}^{*(2)}}{V_{\Gamma}^{*(1)}} = \frac{k+ \gamma (n-1)}{k+\gamma n+ \gamma n^c}.
  \end{equation}
For sufficiently large $c>1$, the ratio $r_{approx}$ will be close to $n^{1-c}$ and arbitrarily small. Thus, if an algorithm could approximate the problem to a factor of $n^{1- c}$ for any $c > 1$, then it could distinguish between the cases of $k+n+n^c$ MDPs with an infinite-horizon return of at least $\gamma R/(1-\gamma)$ and where fewer than $k+n$ MDPs have an infinite-horizon return of at least $\gamma R/(1-\gamma)$ in fMDP-SS. However, this would solve the underlying instance of the \textsc{SetCover} problem, and therefore is impossible unless $P = NP$. Therefore, the fMDP-SS problem is not only NP-hard, but there is no polynomial-time algorithm that can approximate it to within any non-trivial factor (specifically $n^{1- c}$, $c>1$) of the optimal solution. 
\end{proof}

 
 We note that the instance of fMDP-SS constructed in Theorem \ref{thm1} can be directly adapted to a general class of POMDPs (see Remark \ref{remark:pomdp}).  As a result, the same construction and arguments demonstrating inapproximability in the fMDP setting hold for the general class of POMDPs. We formally state this result below. \\


 \begin{corollary}
     The problem of selecting an optimal subset of sensors that can maximize the infinite-horizon expected return provided by the resulting optimal policy for a general class of POMDPs is inapproximable to within a factor of $n^{1-c}$, for any $n, c>1$.
 \end{corollary}

Next, based on similar arguments, we have the following result characterizing the complexity of the fMDP-AS problem. {See Appendix \ref{app:proofs} for complete proof.}\\

\begin{theorem} \label{thm2} Unless $P=NP$, there is no polynomial-time algorithm that can approximate the fMDP-AS Problem to within a factor of $n^{1-c}$, for any $c > 1$.
\end{theorem}

\vspace{-2mm}
\section{Empirical Evaluation of Greedy Algorithm}
Greedy algorithms, which iteratively and myopically choose items that provide the largest immediate benefit, provide computationally tractable and near-optimal solutions to many combinatorial optimization problems \cite{hashemi2020randomized,nemhauser1978analysis}. In this section, we present a greedy algorithm (Algorithm \ref{alg:1}) for the fMDP-SS (resp. fMDP-AS) problem with uniform sensor (resp. actuator) costs. {According to the results presented in the previous section, the greedy algorithm does not admit formal theoretical guarantees in general for the fMDP-SS and fMDP-AS problems. We present some explicit examples where greedy can perform arbitrarily poorly in Appendix \ref{app:failure_greedy}. Despite this, we empirically evaluate the performance of the greedy algorithm in several instances of fMDP-SS and fMDP-AS problems.}


\begin{algorithm}[!htpb]
\caption{Greedy Alg. for fMDP-SS (or fMDP-AS)}
\label{alg:1}
\begin{algorithmic}[1] 

\Require A factored MDP $\mathcal{M}$, set of candidate sensors (or actuators) $X$ with uniform costs, and sensor (or actuator) budget $M$
\Ensure A set of selected sensors (or actuators) $Y$

\State $k \gets 0$, $Y \gets \emptyset$
\While{$k \leq M$}
    \For{$i \in (X \setminus Y)$}
        \State Compute infinite-horizon return $V^*(Y \cup \{i\})$
    \EndFor
    \State $j \gets \arg\max_{i \in X \setminus Y} V^*(Y \cup \{i\})$
    \State $Y \gets Y \cup \{j\}$
    \State $k \gets k + 1$
\EndWhile

\end{algorithmic}
\end{algorithm}

\subsection{Actuator Selection in Electric Networks}
We consider the problem of selecting an optimal subset of actuators which can minimize fault percolation in an electric network, which we will refer to as the Actuator Selection in Electric Networks (ASEN) Problem. We model the distributed micro-grid network as a graph $G = (V,E)$, where each node $i \in V$ represents a micro-grid system and each edge $e_{ij} \in E$ is a link between nodes (or micro-grids) $i$ and $j$ in the electric network (see Figure \ref{fig:intro}(b)). The state of each micro-grid system (or node) $i \in (1, \hdots, |V|)$ is represented by the variable $s_i \in \{ \texttt{healthy}, \texttt{faulty} \}$. If a particular node $i$ is faulty, a neighboring node $j \in \mathcal{N}_i$, where $\mathcal{N}_i$ represents the non-inclusive neighborhood of the node $i$, will become faulty with a certain probability $p_{ji}$. This is known as the \textit{Independent-Cascade} diffusion process in networks, and has been well-studied in the network science literature. Given an initial set of faulty nodes $S \subset V$, as the diffusion/spread process unfolds, more nodes in the network will become faulty. In large-scale distributed micro-grid networks, a central grid controller can perform \textit{active islanding} of micro-grids (or nodes) using islanding switches (or actuators), in order to prevent percolation of faults in the network. Given that only a limited number of islanding switches can be installed, the goal is to determine the optimal placement locations (nodes) in the distributed micro-grid, which can minimize the fault percolation. This problem is similar to the problem of minimizing influence of rumors in social networks studied in \cite{yan2019minimizing}. Similar to \cite{yan2019minimizing}, we wish to identify a blocker set, i.e., a set of nodes which can be disconnected (islanded) from the network, which can minimize the total number of faulty nodes. We assume that the micro-grid system is capable of maintaining power flow and energy balance after isolating certain nodes.

Given a budget $K$, the problem of selecting the optimal subset of $K$ nodes on which actuators, i.e., islanding switches, can be installed, which can maximize the number of healthy nodes in the network (i.e., minimize the spread of faults) is an instance of the fMDP-AS problem presented in this paper. Each node $ i \in (1, \hdots, |V|) $ corresponds to an MDP with state $s_i \in \{ \texttt{healthy}, \texttt{faulty} \}$. The state transition of a node depends on the state transitions of its neighboring nodes and is governed by the influence probabilities $p_{ji}$, for each pair of nodes $(j,i)$. Thus, the overall state of the fMDP can be factored into $n = |V|$ state variables. If a node is in \texttt{healthy} state, it receives a reward of $+1$, and if it is in \texttt{faulty} state, it receives a reward of $0$.  The overall reward of the fMDP at any time step is the sum of the rewards of each node. If an actuator is installed on a particular node $i \in V$, it can perform active islanding to disconnect from the network, and this is denoted by the factored action space $\mathcal{A}_i = \{ \texttt{island}, \texttt{na} \}$. In case of no actuator on a node $i \in V$, it has a default action (do nothing), and this is denoted by the factored action space $\mathcal{A}_i = \{\texttt{na} \}$. The action space of the fMDP is the product of the action spaces of the MDPs corresponding to each node.

We set the discount factor $\gamma=0.95$ and sample $p_{ji} \in (0.2, 0.6)$ uniformly at random for all node pairs $(i,j) \in V \times V$, and compute the infinite-horizon discounted return for the fMDP (electric network) as the fault percolation process unfolds over the network, according to the independent cascade model. We use a greedy influence maximization (IM) algorithm under the independent cascade model to identify the initial set $S$ of faulty nodes with maximum influence for fault propagation. We consider two types of widely used random graph models to generate instances of the network \cite{jelenius2004graph}: (i) Erdős Renyi (ER) random graph denoted by $G(n,p)$, where $n$ is the number of nodes and $p$ represents the edge-probability and (ii) Barabasi-Albert (BA) random graph denoted by $G(n)$, where $n$ is the number of nodes. {We present experiments on  BA network models and additional experiments on ER networks  in Appendix \ref{app:add_exp}.} 

\begin{figure*}[!htpb]
    \centering
    \begin{subfigure}[b]{0.3\textwidth}
    \centering
        \includegraphics[width=140pt]{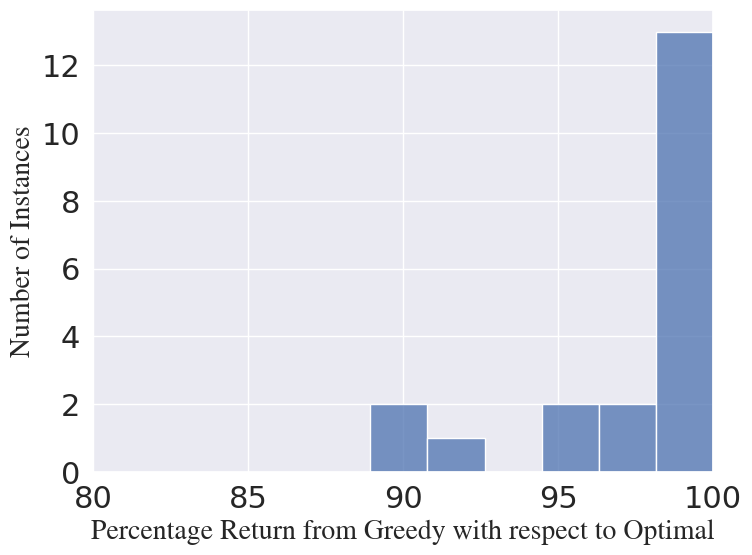}
        \caption{Empirical performance of greedy w.r.t optimal over random network instances of ASEN}
        \label{fig:a}
    \end{subfigure}
    \hfill
        \begin{subfigure}[b]{0.3\textwidth}
    \centering
        \includegraphics[width=130pt]{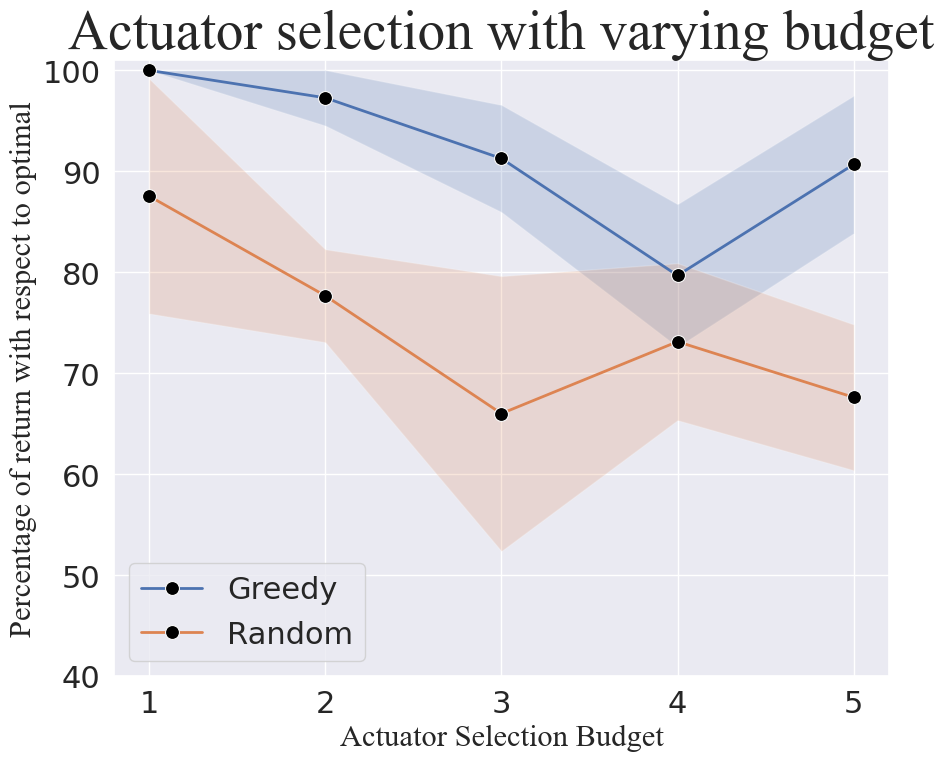}
        \caption{Empirical performance of greedy w.r.t optimal with varying budget}
        \label{fig:b}
    \end{subfigure}
    \hfill
    \begin{subfigure}[b]{0.3\textwidth}
    \centering
        \includegraphics[width=140pt]{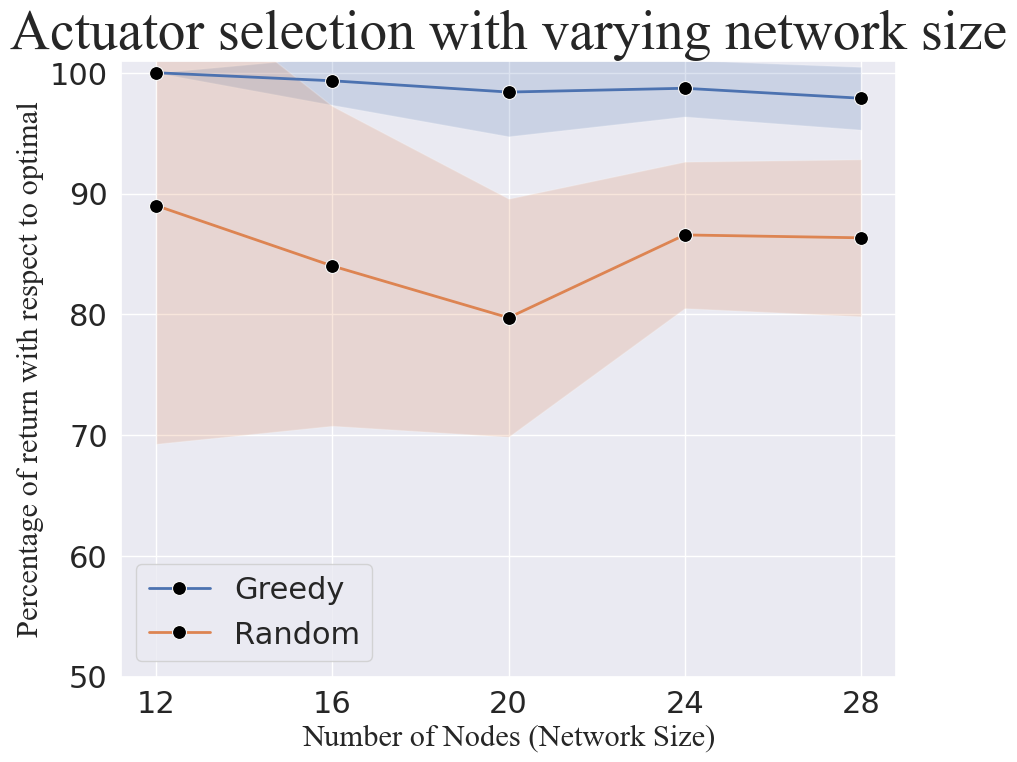}
        \caption{Empirical performance of greedy w.r.t optimal for varying network size}
        \label{fig:c}
    \end{subfigure}
    
    \medskip

    \begin{subfigure}[b]{0.3\textwidth}
    \centering
        \includegraphics[width=135pt]{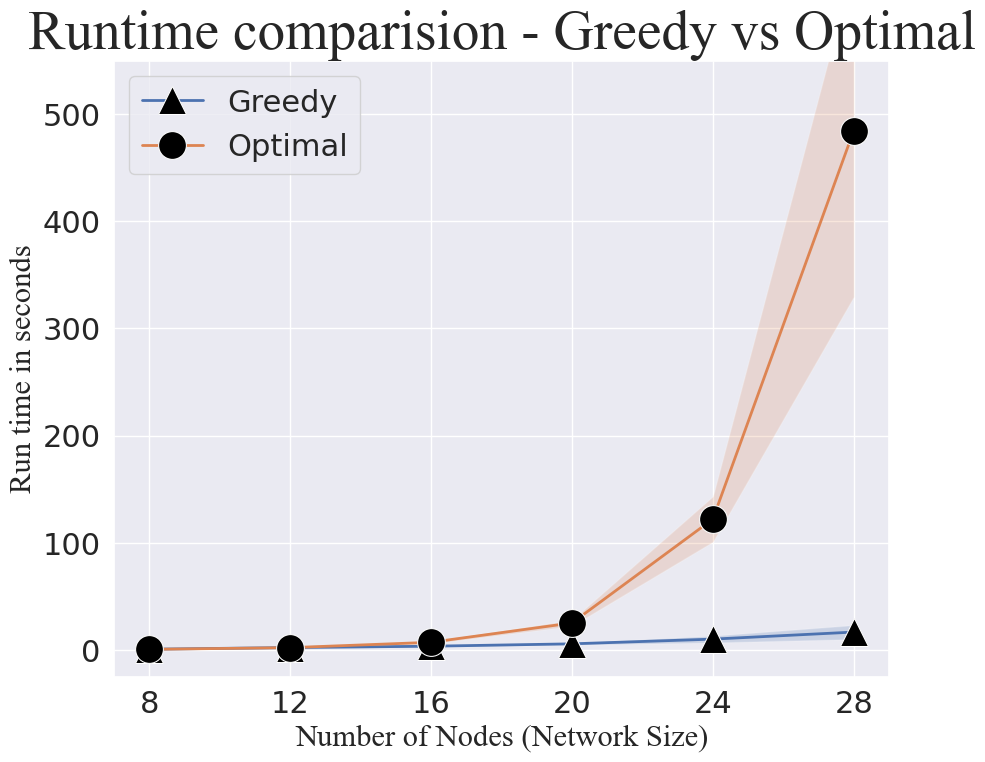}
        \caption{{Runtime comparison of greedy and optimal for varying network size}}
        \label{fig:d}
\end{subfigure}
        \hfill
            \begin{subfigure}[b]{0.3\textwidth}
    \centering
        \includegraphics[width=140pt]{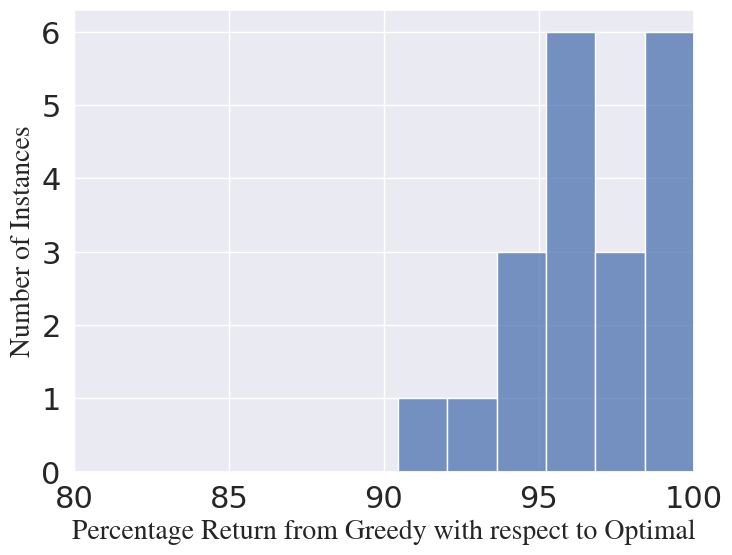}
        \caption{Empirical performance of greedy w.r.t optimal for random instances of fMDP-SS}
        \label{fig:e}
    \end{subfigure}
    \hfill
    \begin{subfigure}[b]{0.3\textwidth}
    \centering
        \includegraphics[width=140pt]{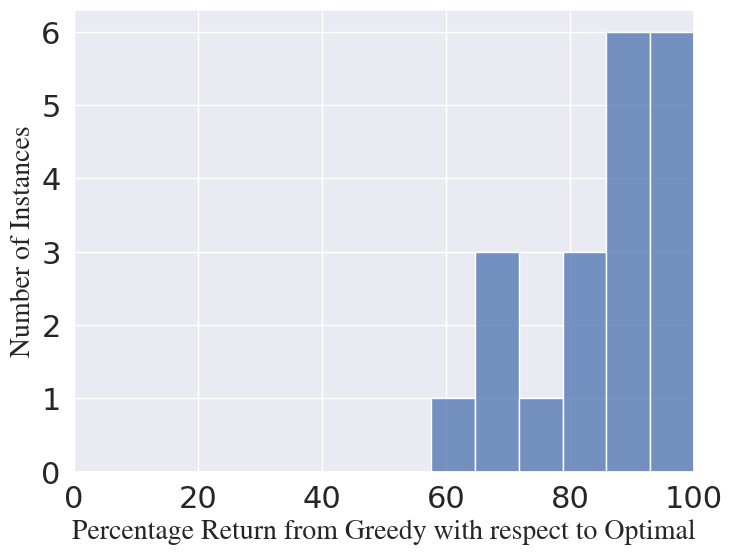}
        \caption{Empirical performance of greedy w.r.t optimal for random instances of fMDP-AS}
        \label{fig:f}
    \end{subfigure}

    \caption{{Empirical evaluation of greedy algorithm for fMDP-SS and fMDP-AS problems. Plots~\ref{fig:a}, \ref{fig:e}, and \ref{fig:f} are histograms that depict the distribution of the greedy algorithm’s performance relative to the optimal solution, capturing both the spread and worst-case performance ratios. Plots~\ref{fig:b}, \ref{fig:c}, and \ref{fig:d} present the average performance ratios/runtimes, along with the corresponding variance around the mean, highlighting the robustness of the greedy algorithm’s performance.}
}
    \label{fig:all}
\end{figure*}

 \textit{Random Instances: } We evaluate the greedy algorithm on the ASEN problem with random ER networks, fixing  $n=30$ with varying $p_{ji}$'s. We set the initial number of faulty nodes to $|S| = 5$ and the selection budget $K=5$ and compare it to optimal (by brute-force search). It can be seen from Figure \ref{fig:a} that greedy shows near-optimal performance  with an average of $96.24\%$ of the optimal.

\textit{Varying Actuator Budget: } We set the initial number of faulty nodes to $|S| = 5$ and run greedy for varying budgets $K \in \{1,2,3,4,5,6,7\}$. The comparison of greedy and random selection with respect to optimal as the selection budget increases is shown in Figures \ref{fig:b}. It can be observed that greedy outperforms random selection. 

\textit{Varying Network Size: }We generate random ER networks with varying network sizes, i.e., $|V| \in  \{8,12,16,20,24,28\}$. We set the initial number of fault nodes to $|S| = 5$ and the actuator selection budget to $K=4$. The comparison of greedy and random selection with respect to optimal as the network size increases is shown in Figures \ref{fig:c}. It can be observed that greedy maintains a near-optimal performance as the network size increases, while the performance of random selection decreases with increase in network size. {We also report the runtimes of the greedy selection and exhaustive optimal subset search in Figure~\ref{fig:d}. As observed, the time required to compute the optimal subset increases rapidly, exhibiting an exponential growth with the network size, becoming infeasible for large networks.}
 
\vspace{-1mm}
\subsection{Evaluation on Random Instances for Robustness} 
 We evaluate the robustness of greedy algorithm over several randomly generated instances of fMDP-SS. Note that any instance of the fMDP-SS problem can be represented as a POMDP to solve for the optimal policy. We use the \texttt{SolvePOMDP} software package \cite{solvepomdp}, a Java program that can solve POMDPs for the optimal policy using incremental pruning combined with state-of-the-art vector pruning methods \cite{walraven2017accelerated}. We run the \textit{exact algorithm} in this package to compute the optimal solution for infinite-horizon cases by setting a value function tolerance $\eta = 1 \times 10^{-6}$ as a stopping criterion.
 We generate 20 instances of fMDP-SS with each instance having $|\mathcal{S}| = 16$ states ($4$ binary state variables) and $|\mathcal{A}| = 16$ actions. The transition function $\mathcal{T}: \mathcal{S} \times \mathcal{A} \times \mathcal{S} \rightarrow [0,1] $ for each starting state  $s \in \mathcal{S} $, action $a \in \mathcal{A} $ and ending state  $s' \in \mathcal{S} $ is a point on a probability simplex $(\Delta_{|\mathcal{S}|})$ is randomly selected. The rewards $\mathcal{R}: \mathcal{S} \times \mathcal{A} \rightarrow \mathbb{R}_{>0} $ for each state-action pair $(s,a)$ are randomly sampled from $\texttt{abs}(\mathcal{N}(0,\sigma))$, with $\sigma \sim \texttt{uniform random} (0,10)$. We consider a set of 4 noise-less sensors $\Omega = \{\omega_1,\omega_2,\omega_3,\omega_4\}$, which can measure the states $(s_1, s_2, s_3, s_4)$, respectively. We consider uniform sensor costs $ c_1 = c_2 = c_3 = c_4 = 1$ and a sensor budget of $C = 2$. We apply a brute-force technique by generating all possible sensor subsets $\Gamma \subset \Omega$ of size $|\Gamma| = 2$, to compute the optimal set of sensors $\Gamma^*$ and compute the optimal return $V_\Gamma^{opt}$ using the solver. We run Algorithm 1 for each of these instances to compute the return $V_\Gamma^{gre}$. It can be seen from Figure \ref{fig:e} that greedy shows near-optimal performance, with an average of $90.19\%$ of the optimal. 

Next, we evaluate the performance of the greedy algorithm over randomly generated instances of the fMDP-AS problem. Note that any instance of the fMDP-AS problem can be solved to find the optimal policy using an MDP solver. We apply the policy iteration algorithm to find the optimal policy and the corresponding infinite-horizon expected return using the \texttt{MDPToolbox} package in Python \cite{mdpsolver}. We generate 20 instances of fMDP-AS with each instance having $|\mathcal{S}| = 20$ states. We consider a set of 10 actuators $\Phi = \{\phi_1, \hdots, \phi_{10}\}$ with uniform selection costs $ k_1 = \hdots = k_{10} = 1$ and a budget of $K = 5$. The action space is given by $\mathcal{A} = \{0,1\} \cup \mathcal{A}_s$, where $\mathcal{A}_s$ is the joint action space corresponding to the selected actuator set. By default, there are two actions available to the agent, i.e., $\{0,1\}$. Each new actuator selected corresponds to a new action variable in the joint action $a$. The transition function $\mathcal{T}: \mathcal{S} \times \mathcal{A} \times \mathcal{S} \rightarrow [0,1] $ for each starting state  $s \in \mathcal{S} $, action $a \in \mathcal{A} $ and ending state  $s' \in \mathcal{S} $ is a value randomly sampled  over a probability simplex $(\Delta_{|\mathcal{S}|})$. The rewards $\mathcal{R}: \mathcal{S} \times \mathcal{A} \rightarrow \mathbb{R}_{>0} $ for each tuple $(s,a)$ are randomly sampled from $\texttt{abs}(\mathcal{N}(0,\sigma))$, with $\sigma \sim \texttt{uniform random} (0,10)$. We apply a brute-force technique to obtain the optimal set of actuators $\Upsilon^*$ and compute the optimal return $V_\Upsilon^{opt}$. We run Algorithm 1 for these instances to compute the return $V_\Upsilon^{gre}$. It can be seen from Figure \ref{fig:f} that greedy shows near-optimal performance, with an average of $85.03\%$ of the optimal.  


\section{Conclusions}

In this paper, we study the design-time sensor and actuator selection problems for  fMDPs under a limited budget, where the objective is to maximize the infinite-horizon return induced by the optimal policy. We show that these problems are not only NP-hard but also admit no polynomial-time approximation algorithm with any non-trivial performance guarantee. These inapproximability results also extend to optimal sensor selection for a broad class of POMDPs. Given this fundamental hardness barrier, we investigate the practical performance of the classical greedy algorithm as a heuristic for sensor/actuator selection. Through extensive experiments on actuator placement in electric grids and on randomly generated instances, we demonstrate that greedy algorithm often achieve near-optimal performance in practice, despite the absence of approximation guarantees. These empirical results highlight the value of greedy algorithms as effective and scalable heuristics for sensor/actuator selection in large-scale fMDPs in practice.

\begin{ack}                        
This material is based upon work supported in part by the Office of Naval Research (ONR) and Saab, Inc. under the Threat and Situational Understanding of Networked Online Machine Intelligence (TSUNOMI) program (grant no. N00014-23-C-1016), the National Science Foundation (NSF) under award 2333487 and the National Aeronautics and Space Administration (NASA) under grant 80NSSC24M0070. Any opinions, findings, and conclusions or recommendations expressed in this material are those of the author(s) and do not necessarily reflect the views of the ONR, Saab, Inc., NSF and/or NASA.  
\end{ack}

\bibliographystyle{IEEEtran}
\bibliography{main.bib}

\appendix

\section{Review of Complexity Theory}
\label{app:complexity_theory}
In this section, we review fundamental concepts from complexity theory \cite{garey1997computers}, which serve as essential background for our theoretical results.\\

\begin{definition}
\label{def:poly_alg}
    A polynomial-time algorithm for a problem is an algorithm that returns a solution to the problem in a polynomial (in the size of the problem) number of computations.
\end{definition}
\vspace{0.2cm}
\begin{definition}
    An optimization problem is a problem whose objective is to maximize or minimize a certain quantity, possibly subject to constraints.
\end{definition}
\vspace{0.2cm}
\begin{definition}
\label{def:dec_ver}
  A decision problem is a problem whose answer is ``yes'' or ``no''. The set P contains those decision problems that can be solved by a polynomial-time algorithm. The set NP contains those decision problems whose ``yes'' answer can be verified using a polynomial-time algorithm.
\end{definition}
\vspace{0.2cm}
\begin{definition}
    A problem $\mathcal{P}_1$ is NP-complete if (a) $\mathcal{P}_1 \in \mathrm{NP}$ and (b) for any problem $\mathcal{P}_2$ in NP, there exists a polynomial time algorithm that converts (or ``reduces'') any instance of $\mathcal{P}_2$ to an instance of $\mathcal{P}_1$ such that the answer to the constructed instance of $\mathcal{P}_1$ provides the answer to the instance of $\mathcal{P}_2$. $\mathcal{P}_1$ is NP-hard if it satisfies (b), but not necessarily (a).
\end{definition}

The \textit{optimization version} of a problem seeks to compute the optimal value of an objective function, such as maximizing reward or minimizing cost. In contrast, the \textit{decision version} asks whether there exists a feasible solution whose objective value exceeds (or falls below) a given threshold. The decision version is crucial for complexity analysis because the class \text{NP} is defined in terms of decision problems. Proving that the decision version of a problem is NP-complete immediately implies that the optimization version is NP-hard. This equivalence allows one to establish computational intractability through polynomial-time reductions, which are well-defined in the decision problem setting.

The above definitions also indicate that if one had a polynomial time algorithm for an NP-complete (or NP-hard) problem, then one could solve every problem in NP in polynomial time. Specifically, suppose we had a polynomial-time algorithm to solve an NP-hard problem $\mathcal{P}_1$. Then, given any problem $\mathcal{P}_2$ in $\mathrm{NP}$, one could first reduce any instance of $\mathcal{P}_2$ to an instance of $\mathcal{P}_1$ in polynomial time (such that the answer to the constructed instance of $\mathcal{P}_1$ provides the answer to the given instance of $\mathcal{P}_2$), and then use the polynomial-time algorithm for $\mathcal{P}_1$ to obtain the answer to $\mathcal{P}_2$. Thus, to show that a given problem $\mathcal{P}_1$ is NP-hard, one simply needs to show that any instance of some other NP-hard (or NP-complete) problem $\mathcal{P}_2$ can be reduced to an instance of $\mathcal{P}_1$ in polynomial time. For NP-hard optimization problems, polynomial-time approximation algorithms are of particular interest.\\

\begin{definition}
An \textit{approximation algorithm} for an optimization problem is an algorithm that always returns a solution whose objective value is within a multiplicative factor $\alpha \in (0,1]$ of the optimal solution. Formally, if $\text{OPT}$ denotes the value of the optimal solution and $\text{ALG}$ denotes the value returned by the algorithm, then for a maximization problem,
\[
\frac{\text{ALG}}{\text{OPT}} \geq \alpha.
\]
\end{definition}

A problem is said to be \textit{inapproximable} within a factor of $\alpha$ if there exists no polynomial-time algorithm that can guarantee a solution satisfying the above inequality for any constant $\alpha > 0$, unless $\text{P} = \text{NP}$. In other words, the approximation ratio $\text{ALG}/\text{OPT}$ can be arbitrarily small, i.e., $\alpha \approx 0$, for some instances of the problem.

In \cite{bhargav2023complexity}, we showed that the problem of selecting sensors for MOMDPs (which are a special class of POMDPs and fMDPs) is NP-hard. In this paper, we show a stronger result that there is no polynomial-time algorithm that can approximate the fMDP-SS (resp., fMDP-AS) Problem to within a factor of $n^{1- c}$ for any $c>1$, even when all the sensors (resp. actuators) have the same selection cost. Specifically, we consider a well-known NP-complete problem, and show how to reduce it to certain instances of fMDP-SS (resp., fMDP-AS) in polynomial time such that hypothetical polynomial-time approximation algorithms for the latter problems can be used to solve the known NP-complete problem.
In particular, inspired by the reductions from Set Cover problem to the influence maximization problems in social networks presented in \cite{kempe2003maximizing} and \cite{schoenebeck2019beyond}, we use the Set Cover problem and provide reductions to the fMDP-SS and fMDP-AS problems in order to establish our inapproximability results. 

\section{Theoretical Proofs}
\label{app:proofs}
\subsection{Proof of Lemma \ref{lma1}}
\begin{proof}
 We will prove both $(i)$ and $(ii)$ as follows. \\
\textbf{Case (i): } Consider the case when state of the fMDP $\Bar{\mathcal{M}}$ is measured using sensor $\omega$. Based on the specified reward function, we can see that the agent can obtain the maximum reward ($R$) at each time-step by taking action $1$ if $s=A$, action $2$ if $s=B$ and any action if $s = C$ or $s=D$. This yields $ V^*(s) = \max_{\pi_\Gamma} V^{\pi_\Gamma}(s) = \sum_{t = 0}^{\infty} \gamma^t R = \frac{ R}{(1-\gamma)}$.  \\
\textbf{Case (ii): } Consider the case when the state of the fMDP $\Bar{\mathcal{M}}$ is not measured (i.e., the sensor $\omega$ is not selected and as a result the agent does not know the current state but only has access to the sequence of actions and rewards).

Due to uncertainty in the state, the agent maintains a belief $b$. The agent performs a Bayesian update of its belief at each time step using the information it has (i.e., the history of actions and observations) \cite{pineau2003point}.  Consider the initial belief $b_0 = [0.5,0.5,0,0]$ for the agent. One can easily verify the following claim: since the agent has an equal probability of being in either state A or state B, it is not optimal for the agent to take either of the actions $a = 1$ or $a = 2$, since they may lead to a large negative reward of $- (1+\delta)R$ by reaching the absorbing state D. The optimal policy is to always take action $a = 0$. Thus, the state of the fMDP will always be either A or B with equal probability and as a result, the belief of the fMDP will always remain $b_t = [0.5,0.5,0,0] $ for all $t>0$. Since taking action $0$ in both state A and B gives a reward of $0$, the expected infinite-horizon reward under the optimal policy is thus
    $V^*(b) = 0$.
\end{proof}



\subsection{Proof of Theorem \ref{thm2}}
Analogous to the proof of the sensor selection case, we first consider a single-state MDP (see Example 2 below) in which installing an actuator enables the optimal policy to obtain a positive return, whereas the absence of an actuator results in zero return (see Lemma \ref{lma2} below). This construction forms the basis for the proof of Theorem \ref{thm2}.
 
\textit{Example 2:} Consider an MDP given by $\Tilde{\mathcal{M}}:= \{\mathcal{S},\mathcal{A},  \mathcal{T}, \mathcal{R}, \gamma \}$. The state space is given by $\mathcal{S} = \{ A, B, C, D\}$. If there is no actuator, the agent can only take the default action $a = 0$, i.e., the action space is $\mathcal{A} = \{ 0\}$ and if there is an actuator installed, then the agent can take one of three actions 0,1, or 2,  i.e., the action space is $\mathcal{A} = \{ 0,1,2\}$. The transition function and reward function are as defined in Example 1. Note that this MDP is similar to the MDP defined in Example 1, but only differs in its action space. We assume that the initial state is $s_0 = A$.\\

\begin{lemma}
 \label{lma2}
For the MDP $\Tilde{\mathcal{M}}$ defined in Example 2, the following holds for any $\gamma\in (0,1)$:
\begin{itemize}
    \item [(i)] If an actuator is installed for $\Tilde{\mathcal{M}}$, the infinite-horizon expected reward under the optimal policy is $V^*(s) = \frac{R}{(1-\gamma)}$ and the optimal policy ensures that the state of the MDP is $s_t=C$, for all $t>0$. 
    \item[(ii)] If no actuator is installed for $\Tilde{\mathcal{M}}$, the infinite-horizon expected reward under the optimal policy is $V^*(s) = 0$ and the optimal policy ensures that the state of the MDP is $s_t\notin \{C,D \}$, for all $t>0$. 
\end{itemize}
\end{lemma} 
\begin{proof}
 We will prove both $(i)$ and $(ii)$ as follows. \\
\textbf{Case (i): } Consider the case when there is an actuator installed on $\Tilde{\mathcal{M}}$. Based on the specified reward function, we can see that the agent can obtain the maximum reward ($R$) at each time-step by taking action $1$ if $s=A$, action $2$ if $s=B$ and any action if $s = C$. This yields $ V^*(s) = \max_{\pi_\Gamma} V^{\pi_\Gamma}(s) = \sum_{t = 0}^{\infty} \gamma^t R = \frac{ R}{(1-\gamma)}$.  \\
\textbf{Case (ii): } Consider the case when there is no actuator installed on $\Bar{\mathcal{M}}$.
 The optimal policy is to always take the only available action $a = 0$ for all $t\geq 0$. Since taking action $0$ in both state A and B gives a reward of $0$, the expected infinite-horizon reward under the optimal policy is thus $V^*(s) = 0$.
\end{proof}

Now, we present the proof for Theorem \ref{thm2}.

\begin{proof} We construct an instance of the fMDP-AS problem using an instance of the \textsc{SetCover} Problem similar to the reduction presented in the proof of Theorem \ref{thm1}, but in this case, each of the MDPs $\mathcal{M}_i$ in the fMDP are defined as in Example 2, except for the reward functions. The effect of selecting a sensor for the $i$'th MDP $\mathcal{M}_i$ as in Example 1 (see Lemma \ref{lma1}) is the same as that of selecting an actuator for the $i$'th MDP $\mathcal{M}_i$ as in Example 2 (see Lemma \ref{lma2}), for $1 \leq i \leq m$. The reward function for each of the MDPs and the overall fMDP are defined as in Theorem \ref{thm1}. 
  
Similar to the arguments made in the proof of Theorem \ref{thm1}, if there is a full set cover $\mathbb{S}_k$ of size $k$ which satisfies $\bigcup_{S_i \in \mathbb{S}_k} S_i  = \mathcal{U}$, then deploying actuators on $k$ MDPs $\mathcal{M}_i$ corresponding sets $S_i$, where $1 \leq i \leq m$, would ensure that the infinite-horizon expected returns of all MDPs $\{\mathcal{M}_{m+1}, \hdots, \mathcal{M}_{N} \}$ will be $\gamma R/(1-\gamma)$. The total infinite horizon expected return in this case is  $V_{\Upsilon}^{*(1)} = (k+\gamma n+ \gamma n^c) R/(1-\gamma)$. Conversely, if there is no full set cover of size $k$ that covers all the elements of the universe $\mathcal{U}$, there are at most $k$ MDPs in $\{\mathcal{M}_{1}, \hdots, \mathcal{M}_{m} \}$ with an infinite-horizon return of $R/(1-\gamma)$ and  fewer than $n$ MDPs in $\{\mathcal{M}_{m+1}, \hdots, \mathcal{M}_{m+n} \}$ with an expected infinite-horizon return of $\gamma R/(1-\gamma)$. The maximum value of the total expected return would be $V_{\Gamma}^{*(2)} = (k+\gamma (n-1)) R/(1-\gamma)$. The approximation ratio $r_{approx}$ is the same as the one obtained in Theorem \ref{thm1} (Equation \ref{eq:badratio}).
  Therefore, it is NP-hard to approximate the fMDP-AS problem to within a factor of $n^{1-c}$ for any $c>1$. 
\end{proof}


\section{Failure of Greedy Algorithm}
\label{app:failure_greedy}

In this section, we provide explicit examples showing how greedy algorithm can perform arbitrarily poorly on some instances of the fMDP-SS and fMDP-AS problems. 
\subsection{Failure of Greedy Algorithm for fMDP-SS}

Consider the following instance of fMDP-SS.

\textit{Example 3:} An fMDP $\mathcal{M}= \{ \mathcal{S}, \mathcal{A}, \mathcal{T}, \mathcal{R}, \gamma, b_0 \}$ consists of 4 MDPs $\{\mathcal{M}_1, \mathcal{M}_2, \mathcal{M}_3, \mathcal{M}_4\}$. Each of these are the single-state variable MDPs as defined in Example 1, except for their reward functions. We will now explicitly define the elements on the tuple $\mathcal{M}$. \\
\textit{State Space $\mathcal{S}$:} The state space of the fMDP $\mathcal{M}$ is the product of the state-spaces of the MDPs $\mathcal{M}_i$, i.e., $\mathcal{S} = \mathcal{S}_1 \times \mathcal{S}_2 \times \mathcal{S}_3 \times \mathcal{S}_4 $. The state of fMDP consists of 4 independently evolving state variables and is given by $s = [s_1,s_2,s_3,s_4]$, where each state variable $s_i$ is as defined in Example 1. We encode the states into a binary representation over $4$ bits as follows: $(A = 1000, B = 0100, C = 0010, D = 0001)$.  \\
\textit{Action Space $\mathcal{A}$:} The action space of the fMDP $\mathcal{M}$ is the product of the action-spaces of the MDPs $\mathcal{M}_i$, i.e.,  $\mathcal{A} = \mathcal{A}_1 \times \mathcal{A}_2 \times \mathcal{A}_3 \times \mathcal{A}_4 $. \\
\textit{Transition Function $\mathcal{T}$:} The probabilistic transition function $\mathcal{T}$ of the fMDP $\mathcal{M}$ is a collection of the individual transition functions $\{ \mathcal{T}_1, \mathcal{T}_2, \mathcal{T}_3, \mathcal{T}_4  \}$ for $s_1,s_2,s_3,s_4$ respectively, where each $\mathcal{T}_i$ is as defined in Example 1.\\
\textit{Reward Function $\mathcal{R}$:} The reward functions of  the MDPs $\{\mathcal{M}_1,\mathcal{M}_2, \mathcal{M}_3 \}$ are independent of each other and are as defined in Example 1 with $R=R_1$ for $\mathcal{M}_1$, $R=R_2$ for $\mathcal{M}_2$ and $R=R_3$ for $\mathcal{M}_3$ respectively. The reward function of $\mathcal{M}_4$ depends on the states $s_2$ and $s_3$ and is defined as:
\begin{equation}
\mathcal{R}_4(s_2,s_3)=
    \begin{cases}
        R_4 & \text{if } s_2 \wedge s_3 = 0010 \\
        0 & \text{otherwise }
    \end{cases}.
\end{equation}

Let $R_4 > R_1 > R_2 > R_3 > 0$ and $R_1 = R_2 + c$ for an arbitrarily small $c$. Let the value of $\delta$ in the individual reward functions for each MDP be such that $\delta > R_4/R_3 -1$. 
The overall reward function for the fMDP $\mathcal{M}$ is given by
\begin{equation}
\label{eq:greedy_value}
\mathcal{R}:= \mathcal{R}_1(s_1,a_1) + \mathcal{R}_2(s_2,a_2) + \mathcal{R}_4(s_2,s_3);
\end{equation}
\textit{Discount Factor $\gamma$:} Let the discount factors of all $\mathcal{M}_i$ be equal to each other and that of $\mathcal{M}$, i.e., $\gamma_1 = \gamma_2 = \gamma_3 = \gamma_4 = \gamma$.

Let $\Omega = \{ \omega_1, \omega_2, \omega_3, \omega_4\}$ be the set of sensors which can measure states, $s_1, s_2,  s_3, s_4$ respectively. Let the cost of the sensors be $\mathcal{C} = (c_1 = 1, c_2 = 1, c_3 = 1, c_4 = 3)$ and the sensor budget be $C=2$. Assume uniform initial beliefs ($b_0$) for all the fMDPs $\mathcal{M}_i$ and $\mathcal{M}$. Since $c_4 > C$,  we apply the greedy algorithm described in Algorithm \ref{alg:1} to this instance of the fMDP-SS with $X = \Omega \setminus \{\omega_4 \}$ and $M=C$. Let the output set $Y = \Gamma$. For any such instance of fMDP-SS, define $r_{gre}(\Gamma) =  \frac{V_{\Gamma}^{gre}}{V_{\Gamma}^{opt}}$,
where $V_{\Gamma}^{gre}$ and $V_{\Gamma}^{opt}$ are the infinite-horizon expected return obtained by the greedy algorithm and the optimal infinite-horizon expected return, respectively. Define $h = R_4/R_2$.\\

\begin{proposition}
\label{prop:1}
For the instance of fMDP-SS problem described in Example 3, the ratio $ r_{gre}(\Gamma)$ satisfies $\lim_{h \rightarrow \infty, c \rightarrow 0} r_{gre}(\Gamma) = 0.$
\end{proposition}
\begin{proof}
We first note that the specified reward function ensures each MDP $\mathcal{M}_i$ follows the optimal policy as in Lemma \ref{lma1} in order to obtain the optimal expected return for the fMDP $\mathcal{M}$. The overall reward of the fMDP $\mathcal{M}$ depends on the individual reward terms $\mathcal{R}_1$, $\mathcal{R}_2$ and $\mathcal{R}_4$. In the first iteration, the greedy algorithm will pick $\omega_1$, because $R_1 > R_2$ by $c$. In the second iteration, greedy would pick $\omega_2$ (because $R_2 > R_3$) and terminate due to the budget constraint. Therefore, the sensor subset selected by the greedy algorithm is $\Gamma = \{ \omega_1, \omega_2 \}$. By Lemma \ref{lma1}, it follows that the infinite-horizon expected reward of the greedy algorithm is 
\begin{equation}
\label{eq:gre}
    V_{\Gamma}^{gre} = \frac{R_1}{1-\gamma} + \frac{R_2}{1-\gamma}  = \frac{2R_2 + c}{1-\gamma}.
\end{equation}
Consider the following selection of sensors for the fMDP-SS instance:  $\Gamma = \{ \omega_2, \omega_3 \}$. By selecting sensors $\omega_2$ and $\omega_3$, the states $s_2 $ and $s_3$ can be measured. By Lemma \ref{lma1}, it follows that the states $s_2$ and $s_3$ are both in $C$ or $(0010)$ at all $t>0$, and thus we have
\begin{equation}
\label{eq:opt}
    V_{\Gamma}^{opt} = \frac{R_2}{1-\gamma} + \frac{R_4}{1-\gamma}  = \frac{R_2 + R_4}{1-\gamma}. 
\end{equation}
 Thus, we have 
\begin{equation*}
\begin{aligned}
    \lim_{h \rightarrow \infty, c \rightarrow 0} & r_{gre}(\Gamma) = \lim_{h \rightarrow \infty, c \rightarrow 0}  \frac{2R_2 + c}{R_2+R_4} =\lim_{h \rightarrow \infty}  \frac{2}{1+h} = 0 
\end{aligned}
\end{equation*} \end{proof}

\begin{remark}
    Proposition \ref{prop:1} means that if we make $R_4$ arbitrarily larger than $R_2$, and $R_1$ slightly larger than $R_2$, the expected return obtained by the greedy algorithm can get arbitrarily small compared to the expected value obtained by the optimal selection of sensors. This is because greedy picks sensors $\omega_1$ and $\omega_2$ due to its myopic behavior. It does not consider the fact that, in spite of $R_3$ being the least reward, selecting $\omega_3$ in combination with $\omega_2$ would eventually lead to the highest reward $R_4$. An expected consequence of the arbitrarily poor performance of the greedy algorithm is that the optimal value function of the fMDP is not necessarily submodular in the set of sensors selected.
\end{remark} 
\subsection{Failure of Greedy Algorithm for fMDP-AS}

Consider the following instance of fMDP-AS.

\textit{Example 4:} An fMDP $\mathcal{M}= \{ \mathcal{S}, \mathcal{A}, \mathcal{T}, \mathcal{R}, \gamma \}$ consists of 4 MDPs $\{\mathcal{M}_1, \mathcal{M}_2, \mathcal{M}_3, \mathcal{M}_4\}$. Each of these are the single-state variable MDPs as defined in Example 2, except for their reward functions. We will now explicitly define the elements on the tuple $\mathcal{M}$ as follows. \\
\textit{State Space $\mathcal{S}$:} The state space of the fMDP $\mathcal{M}$ is the 
same as that in Example 3; \textit{Action Space $\mathcal{A}$:} The action space of the fMDP $\mathcal{M}$ is the product of the action-spaces of the MDPs $\mathcal{M}_i$, i.e., $\mathcal{A} = \mathcal{A}_1 \times \mathcal{A}_2 \times \mathcal{A}_3 \times \mathcal{A}_4 $. The action spaces $\mathcal{A}_i$ depend on the placement of actuators, as detailed in Example 2; \textit{Transition Function $\mathcal{T}$:} The probabilistic transition function $\mathcal{T}$ of the fMDP $\mathcal{M}$ is a collection of the individual transition functions $\{ \mathcal{T}_1, \mathcal{T}_2, \mathcal{T}_3, \mathcal{T}_4  \}$ for $s_1,s_2,s_3,s_4$ respectively, where each $\mathcal{T}_i$ is as defined in Example 2; \textit{Reward Function $\mathcal{R}$:} The reward function of the fMDP $\mathcal{M}$ is the same as that in Example 3 (see Equation \ref{eq:greedy_value}); \textit{Discount Factor $\gamma$:} Let the discount factors of $\mathcal{M}_i$ be equal to the discount factor of $\mathcal{M}$ i.e., $\gamma_1 = \gamma_2 = \gamma_3 = \gamma_4 = \gamma$.

Let $\Phi = \{ \phi_1, \phi_2, \phi_3, \phi_4\}$ be the set of actuators with costs  $\mathcal{K} = (k_1 = 1, k_2 = 1, k_3 = 1, k_4 = 3)$, that can influence the transitions of states, $s_1, s_2,  s_3, s_4$ respectively. Let the actuator selection budget be $K=2$. Since $k_4 > K$, we apply the greedy algorithm described in Algorithm \ref{alg:1} to this instance of the fMDP-AS with $X = \Phi \setminus \{ \phi_4 \}$ and $M=K$. Let the output set $Y = \Upsilon $. For any such instance of fMDP-AS, define $r_{gre}(\Upsilon) =  \frac{V_{\Upsilon}^{gre}}{V_{\Upsilon}^{opt}}$,
where $V_{\Upsilon}^{gre}$ and $V_{\Upsilon}^{opt}$ are the infinite-horizon expected return obtained by the greedy algorithm and the optimal return respectively. Define $h = R_4/R_2$.\\

\begin{proposition}
\label{prop:2}
For the instance of fMDP-AS problem described in Example 4, the ratio $ r_{gre}(\Upsilon)$ satisfies  $\lim_{h \rightarrow \infty, c \rightarrow 0} r_{gre}(\Upsilon) = 0. $

\end{proposition}
\begin{proof}
    The construction of the proof and arguments are similar to Proposition \ref{prop:1}.
\end{proof}

\begin{remark}
    Proposition \ref{prop:2} means that if we make $R_4$ arbitrarily larger than $R_2$, and $R_1$ almost equal to $R_2$, the expected return obtained by greedy can get arbitrarily small compared to the expected value obtained by optimal selection of actuators. This is because greedy picks actuators $\phi_1$ and $\phi_2$ due to its myopic behavior. It does not consider the fact that, in spite of $R_3$ being the least reward, selecting $\phi_3$ would eventually lead to the highest reward $R_4$. An expected consequence of the arbitrarily poor performance of the greedy algorithm is that the optimal value function of the fMDP is not necessarily submodular in the set of actuators selected.
\end{remark}
 The fMDP-SS and fMDP-AS problems are combinatorial optimization problems and the space of possible solutions can be exponentially large. Greedy algorithms, which build a solution incrementally by making a locally optimal choice at each step, might not explore the space effectively. This leads to situations where the greedy algorithm locks into a suboptimal path early in the process, missing out on better solutions that require considering different combinations of sensors. In the counter-example for the fMDP-SS (resp. fMPS-AS) problem we constructed for Proposition \ref{prop:1} (resp. Proposition \ref{prop:2}), a particular sensor's (resp. actuator's) utility gain was not myopically the largest, however, when it was combined with another sensor (resp. actuator), the pair of sensors (resp. actuators) provided a larger increase in utility. This local focus causes the greedy algorithm to miss globally optimal solutions, especially in complex sensor selection problems where interactions between sensors create synergistic effects that are not captured by simply adding one sensor at a time based on immediate gains.

 \section{Additional Experiments}
In this section, we present experiments for the ASEN problem with BA network models and ER network models with fixed diffusion probabilities $(p_{ji})$.
\label{app:add_exp}

\begin{figure*}[!htpb]
    \centering
    \begin{subfigure}[b]{0.3\textwidth}
    \centering
        \includegraphics[width=140pt]{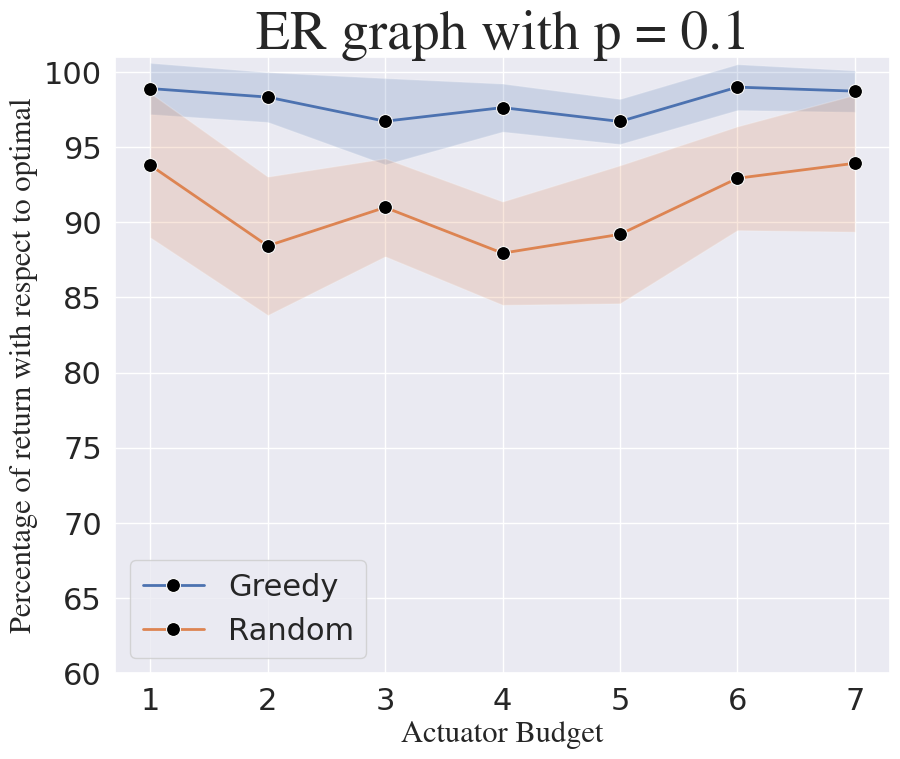}
        \caption{Comparison of greedy v.s. random w.r.t. optimal for ER networks with $p=0.1$ - Varying budget}
        \label{fig:ad}
    \end{subfigure}
    \hfill
    \begin{subfigure}[b]{0.3\textwidth}
    \centering
        \includegraphics[width=140pt]{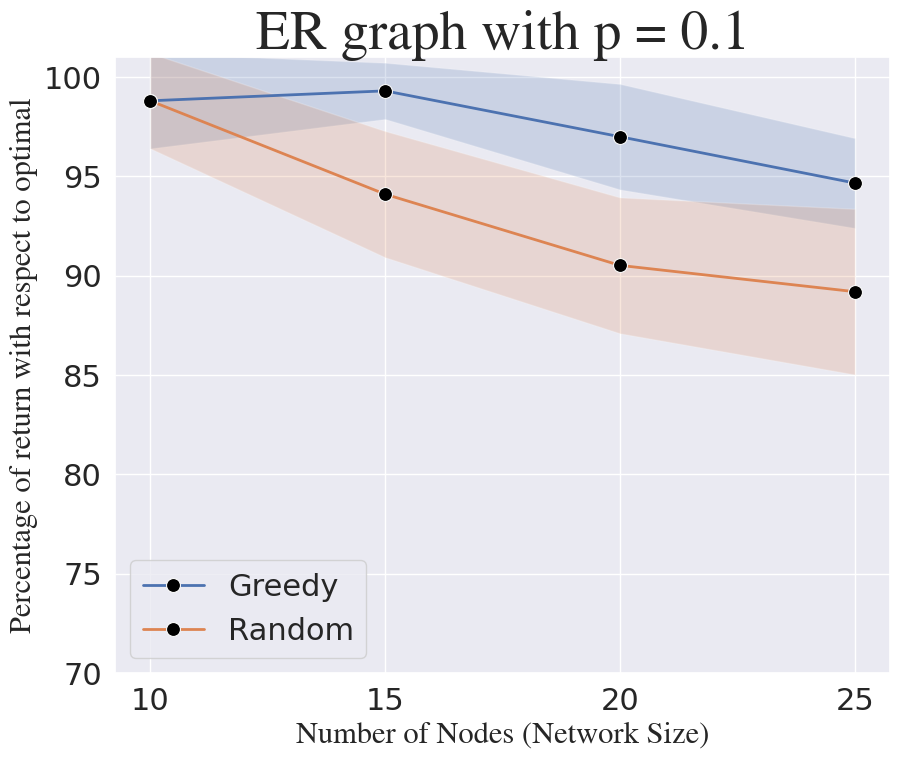}
        \caption{Comparison of greedy v.s. random w.r.t. optimal for ER networks with $p=0.1$ - Varying network size}
        \label{fig:ae}
    \end{subfigure}
    \hfill
    \begin{subfigure}[b]{0.3\textwidth}
    \centering
        \includegraphics[width=140pt]{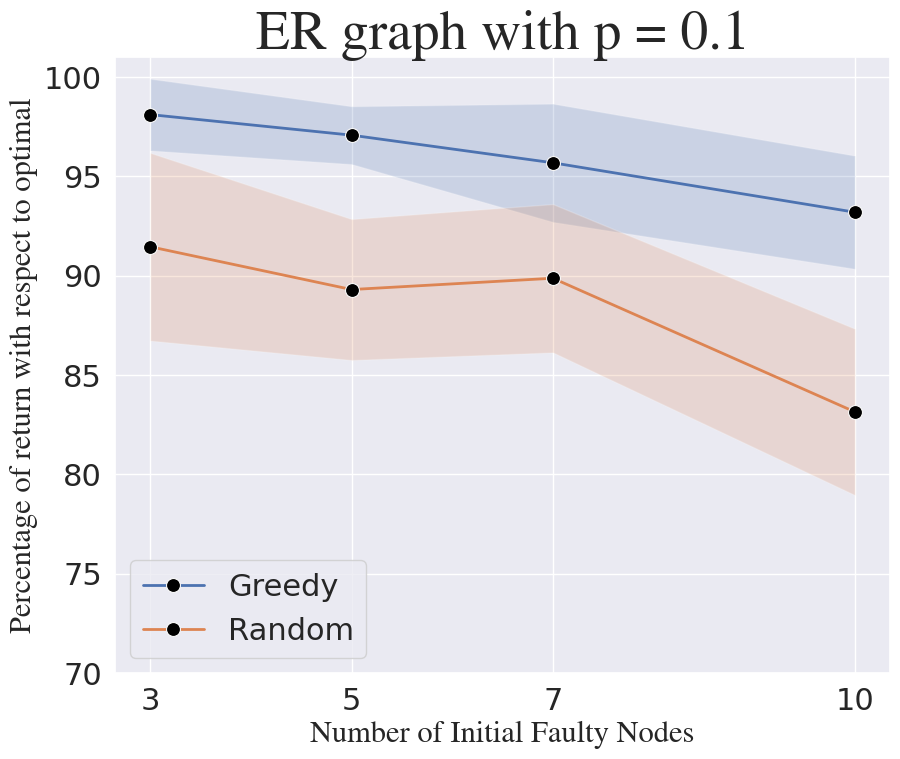}
        \caption{Comparison of greedy v.s. random w.r.t. optimal for ER networks with $p=0.1$ - Varying no. of faulty nodes}
        \label{fig:af}
    \end{subfigure}

    \medskip
    
    \begin{subfigure}[b]{0.3\textwidth}
    \centering
        \includegraphics[width=140pt]{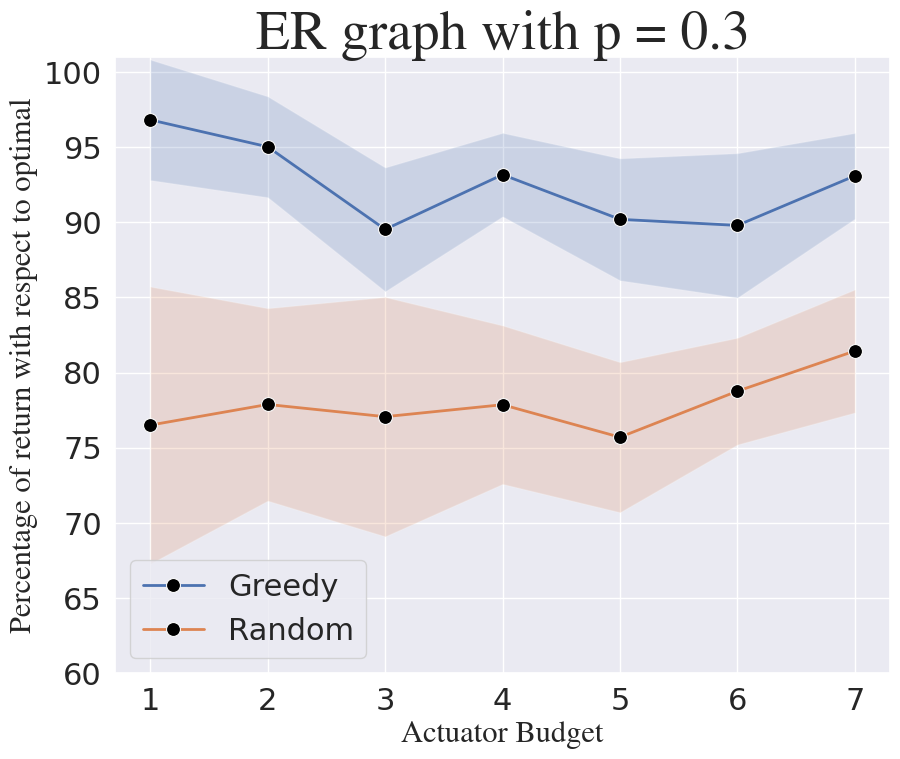}
        \caption{Comparison of greedy v.s. random w.r.t. optimal for ER networks with $p=0.3$ - Varying budget}
        \label{fig:ag}
    \end{subfigure}
    \hfill
    \begin{subfigure}[b]{0.3\textwidth}
    \centering
        \includegraphics[width=140pt]{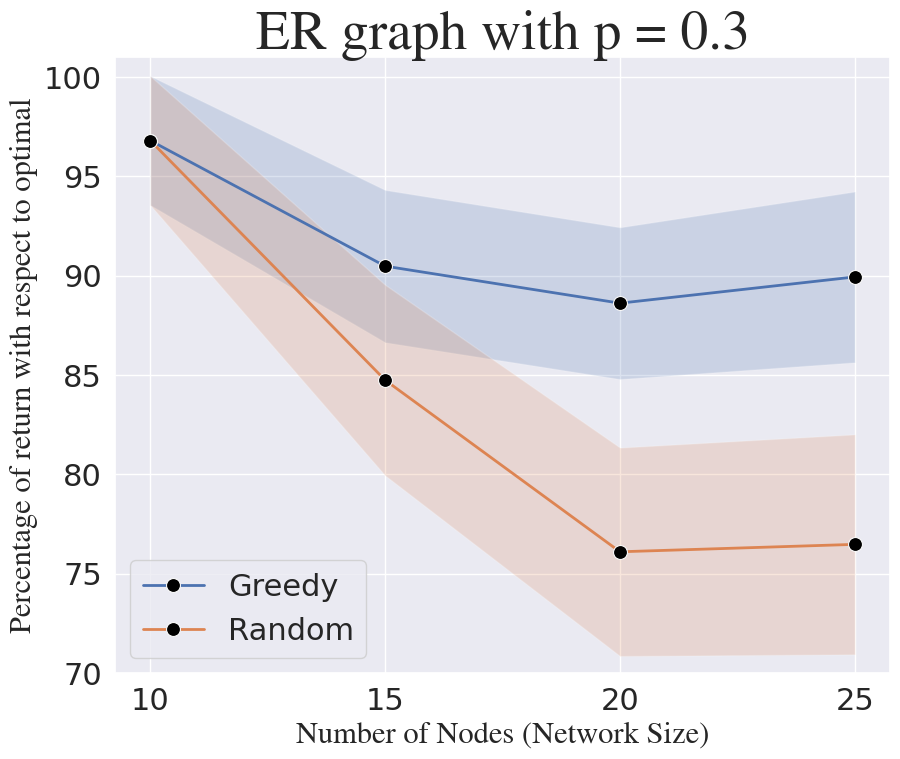}
        \caption{Comparison of greedy v.s. random w.r.t. optimal for ER networks with $p=0.3$ - Varying network size}
        \label{fig:ah}
    \end{subfigure}
    \hfill
    \begin{subfigure}[b]{0.3\textwidth}
    \centering
        \includegraphics[width=140pt]{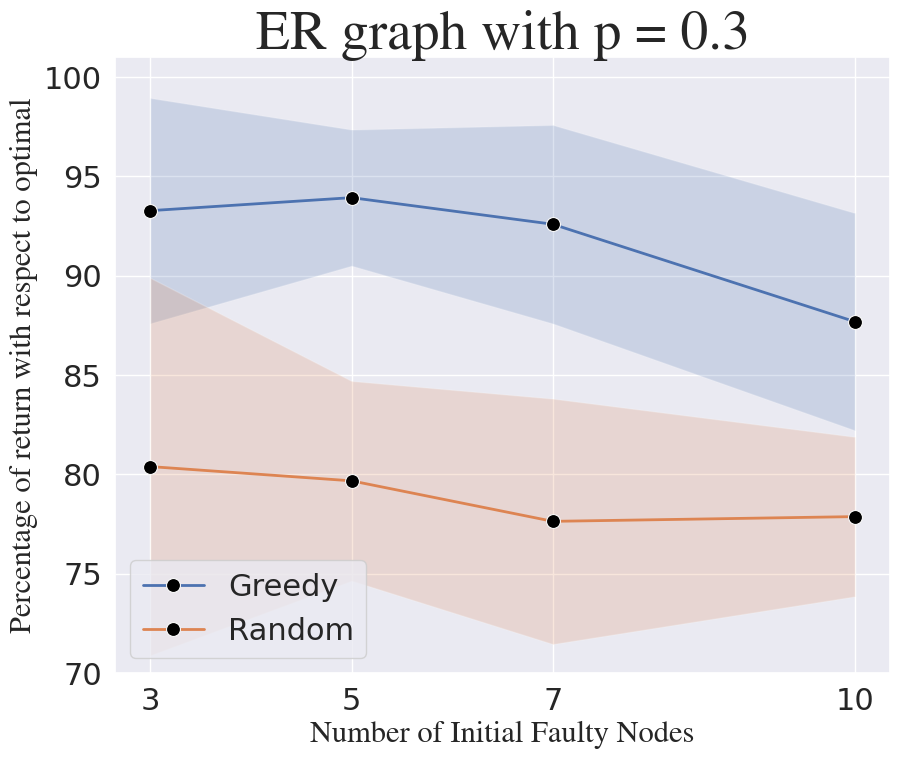}
        \caption{Comparison of greedy v.s. random w.r.t. optimal for ER networks with $p=0.3$ - Varying no. of faulty nodes}
        \label{fig:ai}
    \end{subfigure}

    \medskip
    
    \begin{subfigure}[b]{0.3\textwidth}
    \centering
        \includegraphics[width=140pt]{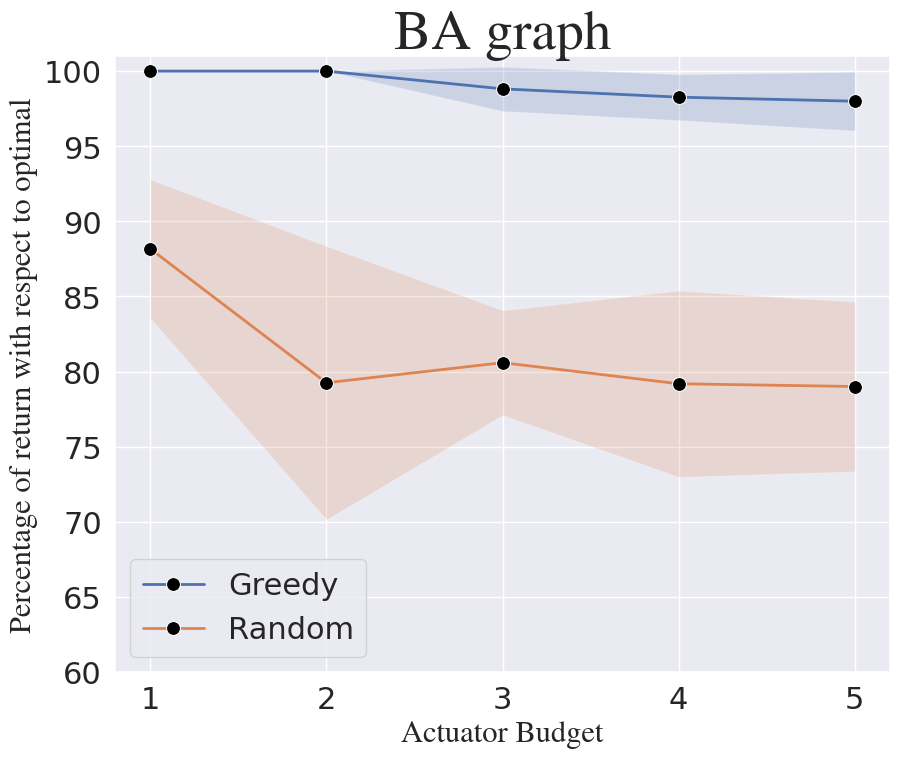}
        \caption{Comparison of greedy v.s. random w.r.t. optimal actuator selection for BA networks - Varying budget}
        \label{fig:aj}
    \end{subfigure}
    \hfill
    \begin{subfigure}[b]{0.3\textwidth}
    \centering
        \includegraphics[width=140pt]{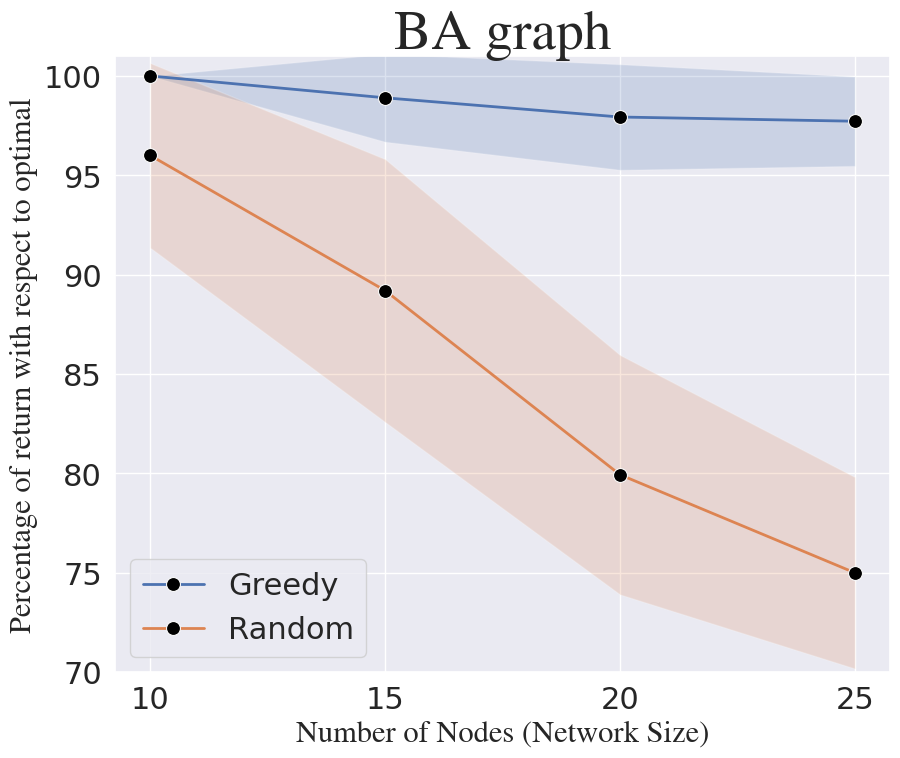}
        \caption{Comparison of greedy v.s. random w.r.t. optimal actuator selection for BA networks - Varying network size}
        \label{fig:ak}
    \end{subfigure}
    \hfill
    \begin{subfigure}[b]{0.3\textwidth}
    \centering
        \includegraphics[width=140pt]{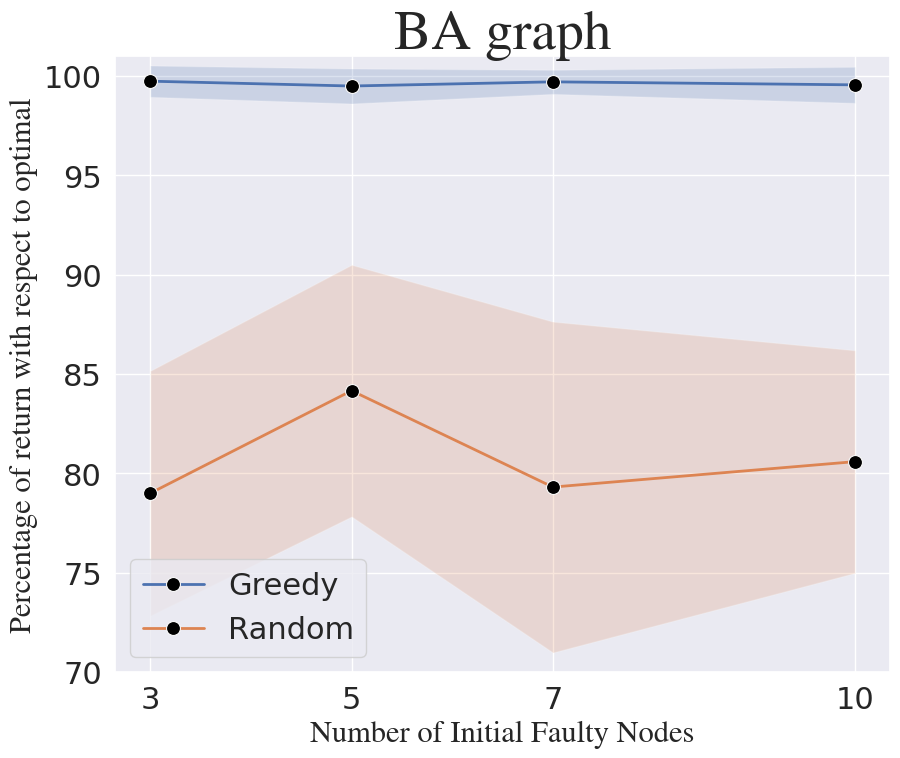}
        \caption*{(l) Comparison of greedy v.s. random w.r.t opt. actuator selection for BA networks - Varying no. of faulty nodes}
        \label{fig:al}
    \end{subfigure}
    \caption{Additional experimental results for evaluating the performance of greedy algorithm for fMDP-SS and fMDP-AS problems}
    \label{fig:app}
\end{figure*}

\textit{Varying Actuator Budget: } We generate random ER networks ($G(30,0.1)$ and $G(30,0.3)$) and BA networks ($G(30)$). We set the initial number of faulty nodes to $|S| = 5$ and run greedy for varying budgets $K \in \{1,2,3,4,5,6,7\}$ for ER networks and $K \in \{1,2,3,4,5\}$ for BA networks. The comparison of greedy and random selection with respect to optimal as the selection budget increases is shown in Figures \ref{fig:app}a, \ref{fig:app}d and \ref{fig:app}g. It can be observed that greedy clearly outperforms random selection for both ER and BA networks, with near-optimal performance. 

\textit{Varying Network Size: } We generate random ER networks ($G(|V|,0.1)$ and $G(|V|,0.3)$) and BA networks ($G(|V|)$) with varying network sizes, i.e., $|V| \in  \{10,15,20,25\}$. We set the initial number of fault nodes to $|S| = 5$ and the actuator selection budget to $K=5$. The comparison of greedy and random selection with respect to optimal as the network size increases is shown in Figures \ref{fig:app}b, \ref{fig:app}e and \ref{fig:app}h. It can be observed that greedy provides near-optimal performance for both ER and BA networks as the network size increases, while the performance of random selection decreases with increase in network size. 

\textit{Varying Number of Faulty Nodes: } We generate random ER networks $(G(30,0.1) \text{ and } G(30,0.3))$ and BA networks ($G(30)$). We vary the initial number of faulty nodes $\lvert S \rvert \in \{ 3,5,7,10 \}$ and set the selection budget to $K = 5$. The comparison of greedy and random selection with respect to optimal as the network size increases is shown in Figures \ref{fig:app}c, \ref{fig:app}f, and \ref{fig:app}l. We observe that greedy provides near optimal performance and consistently outperforms random selection for varying number of faulty nodes in the networks. 

 In all of the above cases, the greedy algorithm provides a performance of more than $70\%$ of the optimal. In particular, the greedy algorithm performs significantly better (with a performance of over $95\%$ of the optimal) for BA networks. 
 

\end{document}